\documentclass[10pt,twoside,a4paper]{amsart}
\usepackage{amsmath}
\usepackage{amsfonts}
\usepackage{amssymb}
\usepackage{amsthm}
\usepackage{newlfont}
\usepackage{graphicx}
\usepackage{amscd}

\theoremstyle{plain}
\newtheorem{Th}{Theorem}
\newtheorem{Cor}[Th]{Corollary}
\newtheorem{Lem}[Th]{Lemma}
\newtheorem{Prop}[Th]{Proposition}
\theoremstyle{definition}
\newtheorem{Def}{Definition}
\theoremstyle{remark}
\newtheorem*{Rem}{Remark}
\numberwithin{equation}{section}
\newcommand{\PP}{{\mathbb P}}
\newcommand{\ZZ}{{\mathbb Z}}
\newcommand{\RR}{{\mathbb R}}
\newcommand{\bx}{\boldsymbol{x}}
\newcommand{\bX}{\boldsymbol{X}}

\begin{document}

\title[B-quadrilateral lattice]
{The B-quadrilateral lattice, its 
transformations \\
and the algebro-geometric construction}

\author[Adam Doliwa]{Adam Doliwa$^\ddagger$}
\thanks{$\ddagger$ Supported in part by the DFG Research Center MATHEON}

\address{Wydzia{\l} Matematyki i Informatyki,
Uniwersytet Warmi\'{n}sko-Mazurski w Olsztynie,
ul. \.{Z}o{\l}nierska 14, 10-561 Olsztyn, Poland}


\email{doliwa@matman.uwm.edu.pl}

%

\begin{abstract}
The B-quadrilateral lattice (BQL) provides geometric interpretation of Miwa's 
discrete BKP equation within the quadrialteral lattice (QL)
theory. After discussing the projective-geometric properties of the lattice we
give the algebro-geometric construction of the BQL 
ephasizing the role of Prym varieties and the corresponding theta functions.
We also present the reduction of the vectorial fundamental transformation of the
QL to the BQL case.
\end{abstract} 
\maketitle

\tableofcontents

\section{Introduction}
During last years the many results describing the well known 
connection between integrable partial
differential equations and the differential geometry of submanifolds have been
transfered to the discrete (difference) level, 
see for example \cite{BobSur,DS-EMP,Schief-JNMP}. The interest in such a
research is stimulated from various fields, like computer visualization,
combinatorics, lattice models in statistical mechanics and quantum field theory,
and recent developments in quantum gravity.

A succesful general approach towards description of the relation between
integrability and geometry is provided by the theory of
multidimensional quadrilateral lattices (QLs)  \cite{MQL}. These are just
maps $x:\ZZ^N\to\PP^M$ ($3\leq N\leq M$) with planar elementary quadrilaterals.
The integrable partial difference equation counterpart of the QLs are
the discrete Darboux equations (see Section \ref{sec:tau} for details), being found 
first \cite{BoKo} as the most general
difference system integrable by the
$\bar\partial$ method. It should be metioned that the (differential)
Darboux equations \cite{Darboux-OS} play
an important role \cite{BoKo-N-KP} in the multicomponent
Kadomtsev--Petviashvilii (KP) hierarchy, 
which is commonly considered \cite{DKJM,KvL}
as the fudamental system of equations in integrability theory.

\begin{figure}[h!]
\begin{center}
\includegraphics{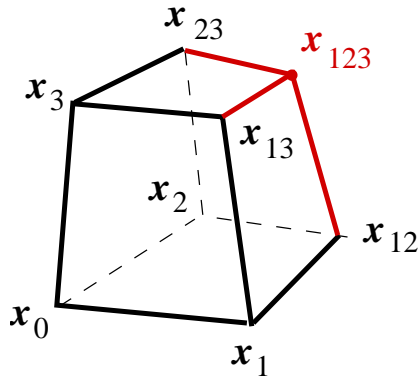}
\end{center}
\caption{The geometric integrability scheme}
\label{fig:TiTjTkx}
\end{figure}
It turns out that integrability of the discrete Darboux system is encoded in a
very simple geometric statement (see Fig.~\ref{fig:TiTjTkx}). 
\begin{Lem}[The geometric integrability scheme] \label{lem:gen-hex}
Consider points $x_0$, $x_1$, $x_2$ and $x_3$ in general position in $\PP^M$,
$M\geq 3$. On
the plane $\langle x_0, x_i, x_j \rangle$, $1\leq i < j \leq 3$ choose a point
$x_{ij}$ not on the lines  $\langle x_0, x_i \rangle$, $\langle x_0,x_j
\rangle$ and $\langle x_i, x_j \rangle$. Then there exists the
unique point $x_{123}$
which belongs simultaneously to the three planes 
$\langle x_3, x_{13}, x_{23} \rangle$,
$\langle x_2, x_{12}, x_{23} \rangle$ and
$\langle x_1, x_{12}, x_{13} \rangle$.
\end{Lem}
Integrable reductions of the quadrilateral lattice (and thus of the discrete
Darboux equations) arise from additional constraints which are compatible with
geometric integrability scheme (see, for example \cite{q-red,DS-sym}). 
One of the most important reductions of the KP hierarchy of nonlinear equations
is the so called BKP hierarchy \cite{DJKM-BKP} 
(here "B" appears in the context of the classification theory of simple 
Lie algebras). In \cite{Miwa} it was shown that the $\tau$-function of the 
BKP hiererchy satisfies certain bilinear discrete equation 
(the transformation between the infinite
sequence of times of the hierarchy and the corresponding discrete variables
is called the Miwa transformation) 
\begin{equation} \label{eq:Hirota-Miwa}
\tau\, \tau_{(123)} = \tau_{(12)}\tau_{(3)} - \tau_{(13)}\tau_{(2)} + 
\tau_{(23)}\tau_{(1)},
\end{equation}
which is known as the discrete BKP or the
Miwa equation.  Here and in all the paper, given a fuction $F$ on 
$\ZZ^N$, we denote its shift in the $i$th direction in a standard manner: 
$F_{(i)}(n_1,\dots, n_i, \dots , n_N) = F(n_1,\dots, n_i + 1, \dots , n_N)$.

The linear problem and the Darboux-type (Moutard)
transformations for the discrete BKP equation \eqref{eq:Hirota-Miwa} were 
constructed in \cite{NiSchief}. In literature there are known several geometric
interpretations of the discrete BKP equation in terms of the reciprocal
figures and inversive geometry~\cite{KoSchiefSBKP,Schief-JNMP}, 
or in terms of the trapezoidal nets \cite{BobSur}. It should be also
mentioned that
the discrete BKP equation has been recently investigated, under the name of the
\emph{cube recurrence}, in combinatorics
\cite{Propp,FominZelevinsky,CarrollSpeyer}.  

In this paper we propose (see Section~\ref{sec:BQL}) another geometric 
interpretation of the discrete BKP
equation, which we consider from the point of view of the quadrilateral lattice 
theory. This new reduction of the quadrilateral lattice, which we call the
\emph{B-quadrilateral lattice} (BQL), is projectively invariant and is based on  
additional local linear constraint. Section~\ref{sec:BQL} is of rather
elementary geometric nature, but the results obtained there have far reaching
consequences. In fact, the paper gives new arguments supporting the 
conjecture that
basic integrability features are consequences of incidence geometry statements 
(see also introductory remarks in \cite{DS-EMP}).

More involved techniques are used in Section~\ref{sec:BQL-Prym}, 
where we elaborate the
algebro-gemetric method to produce large classes of the B-quadrilateral lattices
and the corresponding solutions of the discrete BKP system of equations. 
In doing that we start from recent results of
\cite{DGNS}, where restrictions on the algebro-geometric data of the 
discrete Darboux system \cite{AKV}
compatible with the discussed reduction were given. Then we proceed to 
formulas for the wave and $\tau$-functions of the BQL in terms
of the Prym theta functions related with the algebraic curves
used in the construction. We transfer this way the algebro-geometric method 
of construction of solutions of the BKP hierarchy \cite{DJKM-Prym-BKP} and of its
two-component generalization \cite{VeselovNovikov} to the discrete level.

We also present, in
Section~\ref{sec:B-transf}, the corresponding reduction
of the vectorial fundamental transformation of the quadrilateral lattice
\cite{TQL} and establish its link with the Pfaffian form of the vectorial 
Moutard transformation found in \cite{NiSchief}. 
In Appendix we give alternative proof of a crucial auxilliary
result of the
paper, and we summarize basic properties of Pfaffians.

\section{The B-quadrilateral lattice}
\label{sec:BQL}
We start with discussing the gometric constraint, which imposed on the 
quadrilateral lattice
allows to define its new integrable reduction. Then we proceed to the algebraic
description of such reduced lattice showing its connection with the discrete
BKP equation. Finally, we discuss relation between the 
$\tau$-function of the
B-quadrilateral lattice, and the $\tau$-function of the quadrilateral lattice.
\subsection{Geometric definition of the BQL}
\begin{figure}
\begin{center}
\includegraphics{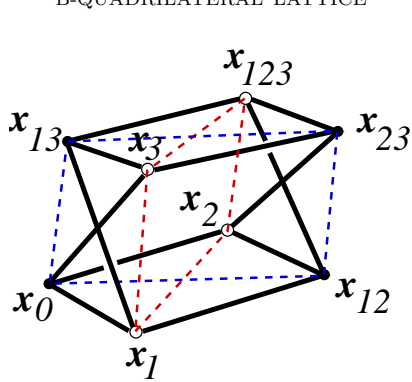}
\end{center}
\caption{Elementary hexahedron of the B-quadrilateral lattice}
\label{fig:moutard}
\end{figure}

\begin{Prop} \label{lem:BKP-hex}
Under hypothesess of Lemma \ref{lem:gen-hex}, assume that the points
$x_0$, $x_{12}$, $x_{13}$, $x_{23}$ are coplanar, then the points
$x_1$, $x_2$, $x_3$, and $x_{123}$ are coplanar as well (see
Figure~\ref{fig:moutard}).
\end{Prop}
It can be shown by the standard linear algebra (for a synthetic-geometry proof
see a remark below). We perform however the calculations, because the way we are
going to do it will be important in next Sections
in showing connection of the BQL with
the discrete BKP equation.
\begin{Lem} \label{lem:BQL-gauge-initial}
Under hypotheses of Proposition \ref{lem:BKP-hex}, for fixed initially
homogeneous coordinates $\bx_0$ and $\bx_1$ (gauges) of $x_0$ and $x_1$,
there exist a gauge 
such that the following linear relations hold
\begin{equation} \label{eq:BQL-gauge-initial}
\bx_{ij} - \bx_0 = f^{ij} (\bx_{i} - \bx_{j}) , \quad 1\leq i< j\leq 3,
\end{equation}
where the coefficients $f^{ij}$ depend on the actual positions of the points
$x_{ij}$.
\end{Lem}
\begin{proof}
The coplanarity of the four points $x_0$, $x_1$, $x_2$ and $x_{12}$ can be
algebraically expressed as the
linear relation 
\begin{equation*}
\alpha\bx_{0} + \beta\bx_{1} + \gamma\bx_{2} + \delta\bx_{12} = 0, 
\end{equation*}
where, by the genericity assumption (no three of the points are collinear), 
all the
coefficients do not vanish. By plaing with rescalling the homogeneous
coordinates of $x_2$ and $x_{12}$ we can transfer above equation 
to the form \eqref{eq:BQL-gauge-initial}
\begin{equation} \label{eq:BQL-gauge-initial-12}
\bx_{12} - \bx_0 = f^{12} (\bx_{1} - \bx_{2}) .
\end{equation}
Similarly, we can rescale the homogeneous
coordinates of $x_3$ and $x_{13}$ to express planarity of the corresponding
elementary quadrilateral as
\begin{equation} \label{eq:BQL-gauge-initial-13}
\bx_{13} - \bx_0 = f^{13} (\bx_{1} - \bx_{3}) .
\end{equation}
However, with fixed gauges $\bx_0$, $\bx_2$ and $\bx_3$ the coplanarity of 
$x_0$, $x_2$, $x_3$ and $x_{23}$ can be
expressed, by plaing with the gauge of $\bx_{23}$, at most as
\begin{equation} \label{eq:gauge-23-bad}
\bx_{23} - \bx_0 = a\bx_{2} - b\bx_{3}. 
\end{equation}
Then
\begin{equation}\label{eq:gauge-23-wedge}
\bx_0\wedge\bx_{12}\wedge\bx_{13}\wedge\bx_{23} = f^{12}f^{13}(a-b)
\bx_0\wedge\bx_{1}\wedge\bx_{2}\wedge\bx_{3},
\end{equation}
and at this moment we use coplanarity of $x_0$, $x_{2}$, $x_{13}$, $x_{23}$, 
which is equivalent to $a=b$.
\end{proof}
\begin{Rem}
Notice that the whole reasoning 
can be applied even if the gauges of $\bx_0$ and of 
$\bx_1$ are not fixed initially (we could then achieve $f^{12}=1$). However we will
need this additional restriction in next Sections.
\end{Rem}
\begin{proof}[Proof of Proposition \ref{lem:BKP-hex}]
By the linear algebra, the homogeneous coordinates of the point 
$x_{123}\in \langle x_3, x_{13}, x_{23} \rangle \cap
\langle x_2, x_{12}, x_{23} \rangle \cap
\langle x_1, x_{12}, x_{13} \rangle$,
in the gauge of Lemma~\ref{lem:BQL-gauge-initial} read
\begin{equation*}
\frac{1}{\rho}\bx_{123} = 
\left( f^{13} - f^{12} + \frac{f^{12}f^{13}}{f^{23}} \right)\bx_1 -
\left(f^{23} + f^{12} - \frac{f^{12}f^{23}}{f^{13}} \right)\bx_2 +
\left( f^{13} - f^{23} + \frac{f^{13}f^{23}}{f^{12}} \right)\bx_3,
\end{equation*}
(we still keep the undetermined yet factor $\rho$)
which gives coplanarity of $x_{123}$, $x_1$, $x_2$ and $x_3$.
\end{proof}
\begin{Cor} \label{cor:x123-x1}
By fixing the gauge function $\rho$ to
\begin{equation*}
\rho = 2 f^{13} - 2 f^{12} - f^{23} + \frac{f^{13}f^{23}}{f^{12}} + 
\frac{f^{12}f^{13}}{f^{23}} + \frac{f^{12}f^{23}}{f^{13}},
\end{equation*}
we find that the
linear relations on the new facets (containing $x_{123}$)
of the cube are of
the form \eqref{eq:BQL-gauge-initial} again, for example
\begin{equation*}
\bx_{123} - \bx_1 = f^{23}_1
(\bx_{12} - \bx_{13}),
\end{equation*}
where 
\begin{equation*}
f^{23}_1 = \frac{f^{23}}{f^{12}f^{13} - f^{12}f^{23} + f^{13}f^{23}}.
\end{equation*}
\end{Cor}
\begin{Rem}
For a geometrically oriented Reader we would like to comment
on another interpretation of Proposition~\ref{lem:BKP-hex}. It
is related to the notion (see, for example \cite{Coxeter})
of the \emph{quadrangular set of points} which are the
intersection points of the lines of a complete quadrilateral (add the
diagonals). Such a configuration is usually denoted by $\mathsf{Q}(ABC,DEF)$, 
where the
first three points $A,B,C$ lie on sides through one vertex while the remaining 
three $D,E,F$ lie on the respectively opposite sides, which form a triangle. 
It is known that $\mathsf{Q}(ABC,DEF)$ implies $\mathsf{Q}(DEF,ABC)$.
\begin{figure}
\begin{center}
\includegraphics{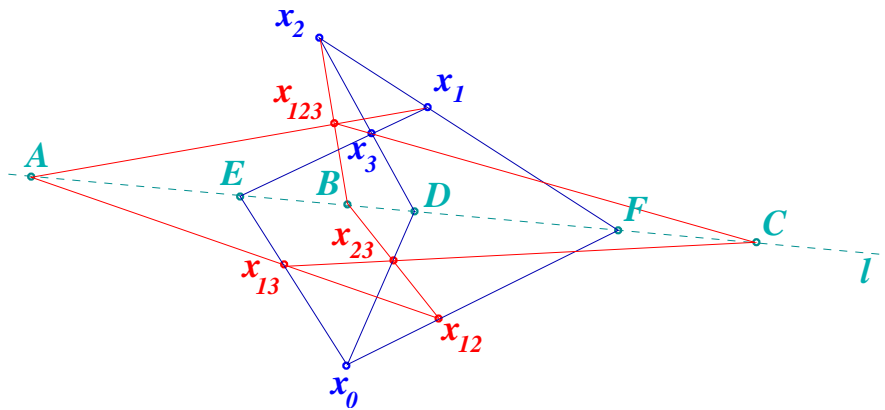}
\end{center}
\caption{Two quadrangles}
\label{fig:quadrangle}
\end{figure}
In notation of Proposition \ref{lem:BKP-hex}, denote by $\ell$ the intersection line
of the plane $\langle x_{1}, x_{2}, x_{3} \rangle $ with the plane 
$\langle x_{12}, x_{23}, x_{13} \rangle $ (containing also
the point $x_{0}$). Denote by $A$, $B$, $C$, $D$, $E$, $F$ intersections of sides 
of the complete quadrilateral with vertices $x_0, x_{12}, x_{23}, x_{13}$ with
$\ell$ (see Figure \ref{fig:quadrangle}), i.e. $\mathsf{Q}(DEF,ABC)$. The
statement of the Lemma is equivalent to the fact that the lines 
$\langle A, x_{1} \rangle $, $\langle B, x_{2} \rangle $, 
$\langle C, x_{3} \rangle $ intersect in one point (which is $x_{123}$); see
Excercise 1 of Section 2.4 of \cite{Coxeter}.
\end{Rem}
\begin{Rem}
As it was pointed to me by Yuri Suris, Proposition \ref{lem:BKP-hex} is
equivalent to the M\"{o}bius theorem \cite{Moebius} on mutually inscribed
tetrahedra. Indeed, vertices $x_0$, $x_1$, $x_2$, $x_3$ of the tetrahedron
$\{x_0, x_1, x_2, x_3 \}$ are contained in the facial planes of the tetrahedron 
$\{x_{12}, x_{13}, x_{23}, x_{123} \}$, and \emph{vice versa}. In fact,
Figure~\ref{fig:quadrangle} appears in M\"{o}bius' original proof of the theorem. 
\end{Rem}
\begin{Rem}
Proposition \ref{lem:BKP-hex} can be also considered as a special version of the 
Miquel theorem \cite{Pedoe} (used in \cite{CDS} to show integrability of the
circular lattice) in the same way 
like Pappus' hexagon theorem is a special case of
the Pascal theorem. 
\end{Rem}

We conclude this Section by defining new reduction of the 
quadrilateral lattice.
\begin{Def} \label{def:BQL}
A quadrilateral lattice $x:\ZZ^N\to\PP^M$ is called the \emph{B-quadrilateral
lattice} if for any triple of different indices $i,j,k$
the points $x$, $x_{(ij)}$,
$x_{(jk)}$ and $x_{(ik)}$ are coplanar.
\end{Def}
\begin{Cor} \label{cor:BKP-impl}
In the B-quadrilateral
lattice, for any triple of different indices $i,j,k$ the points 
$x_{(i)}$, $x_{(j)}$, $x_{(k)}$ and $x_{(ijk)}$ are coplanar.
\end{Cor}

\subsection{Multidimensional consistency of the BQL constraint}
As it was shown in \cite{MQL} the planarity condition, which allows to construct
the point $x_{123}$ as in Lemma \ref{lem:gen-hex}, does not lead to any further
restrictions if we increase dimension of the lattice. This is the consequence of
the of the following geometric observation.
\begin{Lem} \label{lem:4D-consist-QL}
Consider points $x_0$, $x_1$, $x_2$, $x_3$ and $x_4$
in general position in $\PP^M$, $M\geq 4$. Choose generic points 
$x_{ij}\in\langle x_0, x_i, x_j \rangle$, $1\leq i < j \leq 4$,
on the corresponding planes, and using
the planarity condition construct the points 
$x_{ijk}\in\langle x_0, x_i, x_j , x_k\rangle$, $1\leq i < j < k \leq 4$ -- the
remaining vertices of the four (combinatorial) cubes.
Then the intersection point $x_{1234}$ of the three planes 
\[\langle x_{12}, x_{123}, x_{124} \rangle, \;
\langle x_{13}, x_{123}, x_{134} \rangle, \;
\langle x_{14}, x_{124}, x_{134} \rangle \quad \text{in} \quad 
\langle x_{1}, x_{12}, x_{13}, x_{14} \rangle,
\] 
coincides with 
the intersection point of the three planes 
\[\langle x_{12}, x_{123}, x_{124} \rangle, \;
\langle x_{23}, x_{123}, x_{234} \rangle, \;
\langle x_{24}, x_{124}, x_{234} \rangle, \quad \text{in} \quad
\langle x_{2}, x_{12}, x_{23}, x_{24} \rangle,
\] 
which is the same as
the intersection point of the three planes 
\[\langle x_{13}, x_{123}, x_{134} \rangle, \;
\langle x_{23}, x_{123}, x_{234} \rangle, \;
\langle x_{34}, x_{134}, x_{234} \rangle, \quad \text{in} \quad 
\langle x_{3}, x_{13}, x_{23}, x_{34} \rangle,
\] 
and 
the intersection point of the three planes 
\[\langle x_{14}, x_{124}, x_{134} \rangle, \;
\langle x_{24}, x_{124}, x_{234} \rangle, \;
\langle x_{34}, x_{134}, x_{234} \rangle, \quad \text{in} \quad  
\langle x_{4}, x_{14}, x_{24}, x_{34} \rangle.
\]
\end{Lem}
\begin{Rem}
In fact, the point $x_{1234}$ is the unique
intersection point of the four three dimensional subspaces
$\langle x_{1}, x_{12}, x_{13}, x_{14} \rangle$,
$\langle x_{2}, x_{12}, x_{23}, x_{24} \rangle$,
$\langle x_{3}, x_{13}, x_{23}, x_{34} \rangle$,
and
$\langle x_{4}, x_{14}, x_{24}, x_{34} \rangle$ of the four dimensional subspace
$\langle x_{0}, x_{1}, x_{2}, x_{3} , x_{4} \rangle$. This observation
generalizes naturally to the case of more dimensional hypercube with the
planar facets.
\end{Rem}
The goal of this Section is to show an analogous result for B-quadrilateral
lattice. Notice, that in previously known reductions of the quadrilateral
lattice, such as the symmetric \cite{DS-sym} or the quadratic \cite{q-red}
reduction, the additional constraint was
imposed on initial quadrilaterals. Then the multidimensional consistency of the
reduction was the
result of the three dimensional consistency of the constraint and the
multidimensional consistency of the quadrilateral lattice. 

In the BQL case,
the constraint is imposed on the level of elementary cubes. Therefore
its four dimensional consistency
is crucial for integrability of the B-quadrilateral lattice, and once proven, 
implies consistency of the reduction in more dimensions.
\begin{Prop} \label{lem:4D-consist-BQL}
Under hypotheses of Lemma \ref{lem:4D-consist-QL}, assume that the BQL
condition holds for the initial data, i.e., the point $x_0$ belongs to the four
planes $\langle x_{ij}, x_{ik}, x_{jk} \rangle$, $1\leq i<j<k \leq 4$. Then all
the three dimensional (combinatorial) cubes obtained in the construction satisfy
the BQL constaint, i.e.,
\begin{equation*}
x_1 \in \langle x_{123}, x_{124}, x_{134} \rangle, \quad
x_2 \in \langle x_{123}, x_{124}, x_{234} \rangle, \quad
x_3 \in \langle x_{123}, x_{134}, x_{234} \rangle, \quad
x_4 \in \langle x_{124}, x_{134}, x_{234} \rangle.
\end{equation*}
\end{Prop}
\begin{proof}
Consider the gauge of Lemma~\ref{lem:BQL-gauge-initial}.
If we add into the construction points $x_4$ and $x_{14}$, 
then by fixing suitably
their gauges $\bx_4$ and $\bx_{14}$, we
can rewrite the coplanarity condition of $x_0$, $x_1$, $x_4$ and $x_{14}$ 
in the form \eqref{eq:BQL-gauge-initial}.
The same argument like in the proof of Lemma~\ref{lem:BQL-gauge-initial} implies
that the algebraic coplanarity conditions of 
$x_0$, $x_i$, $x_4$ and $x_{i4}$, $i=2,3$ take the form of equation 
\eqref{eq:BQL-gauge-initial}.

By fixing gauges of points $x_{ijk}$ like in Corollary~\ref{cor:x123-x1},
we obtain the relations
\begin{equation} \label{eq:x_ijk-x_i}
\bx_{ijk} - \bx_{i} = f^{jk}_{i} (\bx_{ij} - \bx_{ik} ), \qquad i,j,k
\quad \text{disctinct},
\end{equation}
where
\begin{equation} \label{eq:star-triangle}
f^{jk}_{i} = \frac{f^{jk}}{f^{ij}f^{ik} - f^{ij}f^{jk} + f^{ik}f^{jk}},
\end{equation}
with $f^{ji}=-f^{ij}$.

In equations \eqref{eq:x_ijk-x_i} let us fix $i=1$ and consider the three
pairs $(j,k)$: $(2,3)$, $(2,4)$ and $(3,4)$. Then after simple calculation
we obtain 
the following linear relation
\begin{equation} \label{eq:4D-BQL-x1}
f^{24}_1f^{34}_1 (\bx_{123} - \bx_1) - f^{23}_1 f^{34}_1 (\bx_{124} - \bx_1) +
f^{23}_1 f^{24}_1 (\bx_{134} - \bx_1) =0,
\end{equation} 
which shows that $x_1 \in \langle x_{123}, x_{124}, x_{134} \rangle$. Other
cases are similar. 
\end{proof}
\begin{Cor}
Under assumptions of Proposition \ref{lem:4D-consist-BQL}, the point $x_{1234}$ 
belongs to the
four planes: $\langle x_{12}, x_{13}, x_{14}  \rangle$, 
$\langle x_{12}, x_{23}, x_{24}  \rangle$, 
$\langle x_{13}, x_{23}, x_{34}  \rangle$ and 
$\langle x_{14}, x_{24}, x_{34}  \rangle$.
\end{Cor}
\begin{Rem}
The same procedure can be applied when we increase 
dimension of the hypercube keeping the BQL constraint.
\end{Rem}

\subsection{BQL and the discrete BKP equation}
\begin{Prop}
A quadrilateral lattice $x:\ZZ^N\to\PP^M$ is a B-quadrilateral lattice if and 
only if
it allows for a homogoneous representation $\bx:\ZZ^N\to\RR^{M+1}_{*}$
satisfying the system of discrete Moutard equations (the
discrete BKP linear problem)
\begin{equation} \label{eq:BKP-lin}
\bx_{(ij)} - \bx = f^{ij} (\bx_{(i)} - \bx_{(j)}) , \quad 1\leq i< j\leq N,
\end{equation}
for suitable functions $f^{ij}:\ZZ^N\to\RR$.
\end{Prop}
\begin{proof}
As we have shown above (we present an alternative difference-equation theory
proof in the Appendix), the B-quadrilateral lattices indeed allow for such a
gauge; here the Remark after Lemma~\ref{lem:BQL-gauge-initial} turns out to be
important. Conversely, three equations \eqref{eq:BKP-lin} for the
pairs $(i,j)$, $(i,k)$, $(j,k)$ imply the linear relation  
\begin{equation} \label{eq:M-BQL}
f^{jk}f^{ik}  (\bx_{(ij)} - \bx) + f^{ij}f^{ik} (\bx_{(jk)} - \bx) -
f^{ij}f^{jk}(\bx_{(ik)} - \bx) =0,  \qquad1\leq i < j < k \leq N,
\end{equation} 
expressing coplanarity of the
four points $x$, $x_{(ij)}$, $x_{(jk)}$ and $x_{(ik)}$.
\end{proof}
The system \eqref{eq:BKP-lin} is well known in the literature \cite{NiSchief}. 
Its compatibility leads to the following set of nonlinear equations
\begin{equation} \label{eq:nonl-BQL-f}
1 + f^{jk}_{(i)}(f^{ij} - f^{ik}) = f^{ik}_{(j)} f^{ij}= f^{ij}_{(k)}f^{ik}, 
\qquad i,j,k \quad \text{distinct},
\end{equation}
with $f^{ji}=-f^{ij}$.

\begin{Rem}
The system  \eqref{eq:M-BQL} 
can be actually solved, see for example \cite{BobSur}, 
(we have used this fact in a hidden form already). 
Simply replace the lower index $i$ in the "star-triangle" relation
\eqref{eq:star-triangle} by the shift $(i)$.
\end{Rem}

On the other side, the second equality in the compatibility condition 
implies existence of the potential $\tau:\ZZ^N\to\RR$, 
in terms of which the functions $f^{ij}$ can be written as
\begin{equation} \label{eq:tau}
f^{ij} = \frac{\tau_{(i)}\tau_{(j)}}{\tau \, \tau_{(ij)}}, \qquad i\ne j .
\end{equation}
The first equality can be then rewritten in the form of the 
system of the discrete 
BKP equations \cite{Miwa}
\begin{equation} \label{eq:BKP-nlin}
\tau\, \tau_{(ijk)} = \tau_{(ij)}\tau_{(k)} - \tau_{(ik)}\tau_{(j)} + 
\tau_{(jk)}\tau_{(i)}, \quad 1\leq i< j < k \leq N.
\end{equation}

\begin{Rem}
Two dimensional quadrilateral lattices whose homogeneous coordinates satisfy 
(up to a gauge)
equation \eqref{eq:BKP-lin} are characterized geometrically \cite{DGNS} 
by condition that any point $x$ and its four
second-order neighbours $x_{(\pm 1 \pm 2)}$ are contained in a subspace of
dimension three. Obviously, any two dimensional slide of the B-quadrilateral
lattice fulfills this property, which can therefore serve as definition of a
two dimensional BQL. However the example of
the standard injection $\ZZ^N\to\RR^N\subset\PP^N$ shows that without additional
requirements this property does not characterize completly multidimensional BQL.
\end{Rem}
\subsection{The $\tau$-functions} \label{sec:tau}
In this Section we present relation between the $\tau$-function of the
quadrilateral lattice \cite{DS-sym}, which we donote here by $\tilde{\tau}$, 
and the
above $\tau$-function of the B-quadrilateral lattice. From the relation 
between 
the KP and BKP hierarchies \cite{DKJM} we expect that 
within the class of the
B-quadrilateral lattices $\tilde{\tau}$ should be equal to the square of $\tau$.

Let us recall briefly the algebraic construction of the $\tau$-function of the
quadrilateral lattice (the
geometric meaning is presented in \cite{DS-sym}). 
The non-homogeneous coordinates $\mathrm{\bx}:\ZZ^N\to\RR^M$
(we restrict our attention to the affine
geometric aspects of the theory) of the quadrilateral lattice satisfy
the system of Laplace equations
\begin{equation} \label{eq:Laplace-a}
\bx_{(ij)} - \bx = a^{ij}(\bx_{(i)} - \bx) + a^{ji}(\bx_{(j)}-\bx), \quad i\ne
j.
\end{equation}
The functions $a^{ij}:\ZZ^N \to\RR$ are not arbitrary (the system
\eqref{eq:Laplace-a} must be compatible), in particular they
can be parametrized in terms of the potentials $h^i$ (the Lam\'{e} coefficients)
as follows
\begin{equation} \label{eq:a-h}
a^{ij} = \frac{h_{i(j)}}{h_i}, \quad i\ne j.
\end{equation}
Define the so called rotation coefficients $\beta_{ij}$ from equations
\begin{equation} \label{eq:lin-h}
\Delta_i h_j = h_{i(j)}\beta_{ij}, \quad i\ne j,
\end{equation}
and the normalized tangent vectors $\bX_i$ from
\begin{equation} \label{eq:x-hX}
\Delta_i \bx = h_i\bX_i .
\end{equation}
Then the Laplace system \eqref{eq:Laplace-a} takes the first
order form
\begin{equation} \label{eq:lin-X}
\Delta_j \bX_i = \beta_{ij}\bX_j ,\quad i\ne j,
\end{equation}
and its compatibility reads
\begin{equation} \label{eq:nlin-beta}
\Delta_j\beta_{ik} = \beta_{ij(k)}\beta_{jk}, \quad i,j,k \quad
\text{distinct}.
\end{equation}
The discrete Darboux equations \eqref{eq:nlin-beta} imply existence of the
potential $\tilde\tau$ (the $\tau$-function of the quadrilateral lattice)
\begin{equation} \label{eq:tau-QL}
\frac{\tilde{\tau}\tilde{\tau}_{(ij)}}{\tilde{\tau}_{(i)}\tilde{\tau}_{(j)}} =
1 - \beta_{ij}\beta_{ji},
\quad i\ne j.
\end{equation}

In looking for the Lam\'{e} coefficients $h_i$ in the reduction from QL to BQL 
we can 
compare both linear systems
\eqref{eq:BKP-lin}  and \eqref{eq:Laplace-a}, and the 
expresions \eqref{eq:a-h} and \eqref{eq:tau} to
obtain
\begin{equation} \label{eq:h-tau}
h_i = (-1)^{\sum_{k<i}m_k}\frac{\tau}{\tau_{(i)}}.
\end{equation}
The corresponding rotation coefficients are then given by (below we assume $i<j$)
\begin{align} \label{eq:beta-BQL-ij}
\beta_{ij} & = - (-1)^{\sum_{i\leq k<j}m_k}
\left( \frac{\tau_{(i)}}{\tau} + \frac{\tau_{(ij)}}{\tau_{(j)}}\right) 
\frac{\tau}{\tau_{(j)}},\\
\label{eq:beta-BQL-ji}
\beta_{ji} & = - (-1)^{\sum_{i\leq k<j}m_k}
\left( \frac{\tau_{(j)}}{\tau} - \frac{\tau_{(ij)}}{\tau_{(i)}}\right)
\frac{\tau}{\tau_{(i)}},
\end{align}
which implies (compare with formula \eqref{eq:tau-QL})
\begin{equation} 
1 - \beta_{ij}\beta_{ji} = \left( 
\frac{{\tau}{\tau}_{(ij)}}{{\tau}_{(i)}{\tau}_{(j)}} \right)^2,
\qquad i\ne j.
\end{equation}
Therefore, we can summarize the above considerations as follows.
\begin{Prop}
Given B-quadrilateral lattice $x$ with the $\tau$-function $\tau$. Then,
formally on the level of the reduction of the system of
discrete affine Laplace equations
\eqref{eq:Laplace-a} to the system of discrete Moutard equations 
\eqref{eq:BKP-lin}, the Lam\'{e} functions and the rotation coefficients 
are given in terms of $\tau$ by equations 
\eqref{eq:h-tau}-\eqref{eq:beta-BQL-ji}, and the corresponding $\tau$-function  
$\tilde{\tau}$ of the quadrilateral lattice is given as
\begin{equation} \label{eq:tau-QL-BQL}
\tilde{\tau} = \tau^2.
\end{equation}
\end{Prop}
\begin{Rem}
We should be aware that the $\tau$-function of the quadrilateral lattice 
is defined with
respect to the coefficients of the affine Laplace equation. Equation 
\eqref{eq:BKP-lin},
although formally written in the affine form, is a consequence of the
projectively-invariant definition of the B-quadrilateral lattice. 
Therefore, in this formal correspondence the
geometric meaning of the rotation coefficients
in the BQL reduction has been lost. We mention that the
affine geometric meaning of equation 
\eqref{eq:BKP-lin}, and therefore also of the rotation coefficients
\eqref{eq:beta-BQL-ij}-\eqref{eq:beta-BQL-ji}, can be 
provided  within the context of the trapezoidal lattices 
\cite{BobSur}
(see also \cite{Sauer,KoSch-trap}).
\end{Rem}
\section{Algebro-geometric construction of the BQL} \label{sec:BQL-Prym}
Below we apply the algebro-geometric approach, well known in the theory if
integrable systems \cite{BBEIM}, to the B-quadrilateral lattice reduction.
It is known \cite{DGNS} that in the BQL
reduction case, the generic algebraic curve, used to generate solutions of the
discrete Darboux equations \cite{AKV}, should be replaced by a curve admitting 
a holomorphic involution with two fixed points. Such curves have been already
used in construction of solutions of the BKP hierarchy \cite{DJKM-Prym-BKP} 
and of its two-component generalization \cite{VeselovNovikov}. We develop the
corresponding results of \cite{DGNS} and we present the explicit formulas for
the lattice points and the solutions of the discrete BKP equation in terms of
the Prym theta functions related to such special curves.

\subsection{Curves with involution and their Prym varieties}
Let us first summarize some fact from theory of Riemann surfaces (see
\cite{FarkasKra,Fay}).
Consider $\hat{\Gamma} \stackrel{\pi}{\to}\Gamma$ a ramified double 
covering of genus $\hat{g}=2g$ of a compact Riemann surface 
$\Gamma$ of 
genus $g$ with exactly two branch points $Q_0$, $Q_\infty$. 
Denote by $\sigma:\hat{\Gamma}\to\hat{\Gamma}$ the holomorphic involution 
permuting
sheets of the covering, i.e., $\Gamma =\hat{\Gamma}/\sigma$. 

The map $\pi^*:\mathcal{J}({\Gamma})\to\mathcal{J}({\hat{\Gamma}})$ lifting
divisor classes of degree $0$ is an injection.
The holomorphic involution $\sigma$ extends to 
$\mathcal{J}({\hat{\Gamma}})$ and allows to define the Prym variety 
\begin{equation} \label{def:Prym-variety}
\mathcal{P}_\sigma({\hat{\Gamma}}) = \{ A- \sigma(A) \, |A\in 
\mathcal{J}(\hat{\Gamma}) \}.
\end{equation}
The natural epimorphism 
$i:\mathcal{J}({\Gamma})\times \mathcal{P}_\sigma({\hat{\Gamma}})
\to\mathcal{J}({\hat{\Gamma}})$ has finite kernel consisting of $4^g$
half-periods in $\mathcal{J}({\Gamma})$.

There exists a basis of cycles $a_i,b_i$, $1\leq i \leq 2g$ on $\hat{\Gamma}$
with the canonical intersection matrix
such that $\pi(a_k),\pi(b_k)$, $1\leq k \leq g$, is a canonical basis of cycles on
$\Gamma$ and 
\begin{equation*}
\sigma(a_k) = - a_{g+k}, \quad \sigma(b_k) = - b_{g+k} , 
\qquad 1\leq k \leq g .
\end{equation*}
The corresponding normalized holomorphic differentials $\omega_i$ 
\begin{equation} \label{eq:a}
\oint_{a_j}\omega_i = \delta_{ij}, \qquad 1\leq i,j \leq 2g,
\end{equation}
satisfy
\begin{equation}
\sigma^*(\omega_k) =-\omega_{g+k},\quad
\sigma^*(\omega_{g+k})  =-\omega_{k}, \qquad 1\leq k \leq g.
\end{equation}
The differentials
\begin{equation*}
u_k = \omega_k - \omega_{g+k}, \qquad \sigma^*(u_k) = u_k, 
\qquad 1\leq k \leq g,
\end{equation*}
form a basis of normalized holomorphic differentials on 
$\Gamma$, while the 
odd differentials
\begin{equation} \label{eq:w}
w_k = \omega_k + \omega_{g+k}, \qquad \sigma^*(w_k) = - w_k, 
\qquad 1\leq k \leq g,
\end{equation}
are called normalized holomorphic
Prym differentials. Then the Riemann matrix 
\begin{equation}
\hat{B}_{jk}  = \oint_{b_j}\omega_k, \qquad \qquad 1\leq j,k \leq \hat{g}
\end{equation}
for 
$\hat{\Gamma}$
has the form
\begin{equation}
\hat{B} = \frac{1}{2}\left( \begin{array}{cc}
\Pi + B & \Pi - B \\
\Pi - B & \Pi + B
\end{array} \right),
\end{equation}
where 
\begin{equation}
B_{jk}  = \oint_{b_j}u_k, \qquad \qquad 1\leq j,k \leq g
\end{equation}
is the corresponding Riemann matrix for $\Gamma$,
and $\Pi$ is the matrix of the $b$-periods of the Prym differentials
\begin{equation} \label{eq:Pi}
\Pi_{jk}  = \oint_{b_j}w_k, \qquad \qquad 1\leq j,k \leq g.
\end{equation}
The matrix $\Pi$ is symmetric and has positively defined imaginary part, and
defines the Prym theta function $\theta(\mathbf{z} ;\Pi)$,
$\mathbf{z}\in\mathbb{C}^{g}$, 
\begin{equation}
\theta(\mathbf{z};\Pi) = \sum_{\mathbf{n}\in{\ZZ^{{g}}}} 
\exp \left\{ \pi i \langle \mathbf{n},{\Pi} \mathbf{n} \rangle + 
2 \pi i\langle  \mathbf{n}, \mathbf{z} \rangle \right\},
\end{equation}
where $\langle \cdot , \cdot \rangle$ denotes the standard bilinear
form in $\mathbb{C}^{{g}}$.

With the above choice of the period matrix $B$ for $\hat{\Gamma}$ 
and with $Q_\infty$ as the base-point of the Abel map 
\begin{equation}
\mathbf{A}(P)=\int_{Q_\infty}^P \boldsymbol{\omega},
\qquad \boldsymbol{\omega}=(\omega_1,\dots ,\omega_{2g}),
\end{equation}
the lift of $\sigma$ from 
$\mathcal{J}(\hat{\Gamma})=\mathbb{C}^{2g}/(I_{2g},\hat{B})$ to 
$\mathbb{C}^{2g}$ reads
\begin{equation*}
\sigma( z_1, \dots , z_{2g}) = 
-(z_{g +1},\dots , z_{2g}, z_1,\dots ,z_{g}),
\end{equation*}
while the map $\pi^*:\mathcal{J}(\Gamma)\to\mathcal{J}(\hat{\Gamma})$ 
is represented by
\begin{equation*}
\pi^*(z_1,\dots ,z_{g}) = (z_1,\dots ,z_{g},-z_1,\dots ,-z_{g}).
\end{equation*}
The Prym variety is a principally polarized abelian variety isomorphic to
$\mathcal{P}=\mathbb{C}^{g}/(I_{g},\Pi)$, and the injection 
$\phi: \mathbb{C}^{g}/(I_{g},\Pi) \to \mathbb{C}^{2g}/(I_{2g},\hat{B})$
is given by
\begin{equation*}
\phi(z_1,\dots ,z_{g}) = (z_1,\dots ,z_{g},z_1,\dots ,z_{g}) .
\end{equation*}
There holds the following analog of the Riemann theorem.
\begin{Lem}[\cite{Fay}] \label{lem:R-P}
Given $\mathbf{e} \in \mathbb{C}^{g}$ such that
$\theta\left(\mathbf{e};\Pi\right) \neq 0$, then the zero divisor $Z$ of
$\theta\left(\int_{Q_\infty}^P\boldsymbol{w} - \mathbf{e};\Pi\right)$ on 
$\hat{\Gamma}$
is of degree $2g$ and satisfies the relation
\begin{equation} \label{eq:R-P}
\phi(\mathbf{e}) =
\mathbf{A}(Z)-\mathbf{A}( Q_0 ) + \pi^* \mathbf{K}_\Gamma, 
\qquad \mathrm{mod} \; (I_{2g},\hat{B}),
\end{equation}
where $\mathbf{K}_\Gamma$ is the Riemann constants vector of $\Gamma$.
Moreover $Z + \sigma(Z) -Q_0 - Q_\infty$ is a canonical divisor on 
$\hat{\Gamma}$.
\end{Lem}
By $\omega_{ST}$, ($S,T\in \hat{\Gamma}$, $S\neq T$) denote the unique 
meromorphic
differential holomorphic in $\hat{\Gamma}\setminus\{S,T\}$ with 
poles of the first
order in $S$, $T$ with residues, correspondingly, $1$ and $-1$, 
and normalized by conditions  
\begin{equation} \label{eq:omST-a}
\oint_{a_j}\omega_{ST} =  0, \qquad 1\leq j \leq 2g.
\end{equation}
It is known that the $b$-periods of such differentials are given by
\begin{equation} \label{eq:omST-b}
\oint_{b_j}\omega_{ST} =  2\pi i \int_{T}^S \omega_j, \qquad 1\leq j \leq 2g,
\end{equation}
with the integral being taken along a curve joining $S$ to $T$ in
$\hat{\Gamma}\setminus \bigcup_{j=1}^{2g}a_j \setminus \bigcup_{j=1}^{2g}b_j$.
Moreover, the following relation between two such differentials holds 
(with the paths of integration being appropriately choosen \cite{FarkasKra})
\begin{equation}\label{eq:omST-PQ}
\int_{P}^Q \omega_{ST} = \int_{S}^T \omega_{PQ}.
\end{equation} 
\begin{Rem}
Such a form can be expressed in terms of the theta function on $\hat{\Gamma}$ 
as
\begin{equation*}
\omega_{ST} = d_P\log\frac{\theta(\mathbf{A}(P) - \mathbf{A}(S) - 
\boldsymbol{\xi};\hat{B})}
{\theta(\mathbf{A}(P) - \mathbf{A}(T) - \boldsymbol{\xi};\hat{B}))},
\end{equation*}
where $\boldsymbol{\xi}$ is a general point of the divisor $\Theta$ of zeros of 
the theta function in
$\mathcal{J}(\hat{\Gamma})$.
\end{Rem}
\subsection{Explicit algebro-geometric formulas}
Given $N$ points $Q_i\in\Gamma$, different from $Q_0, Q_\infty$, 
and an effective
non-special divisor ${D}$ of degree $\hat{g}$ such that 
\begin{equation} \label{eq:D-C}
D + \sigma(D) - Q_0 - Q_\infty \stackrel{\mathcal{J}_{2\hat{g}-2}({\hat{\Gamma}})}{=} 
C_{\hat{\Gamma}}
\end{equation}
is a canonical divisor. For an arbitrary $m\in\ZZ^N$ there exists \cite{DGNS}
the unique function
$\psi(m)$ meromorphic on $\Gamma$ having in points $Q_i$ (in points
$\sigma(Q_i)$) poles
(corrspondingly, zeros) of the order $m_i$, no other singularities except for possible 
simple poles in points of the divisor $D$, and normalized to $1$ at $Q_\infty$. 
In \cite{DGNS} it was shown that, as a function of the discrete parameter $m$,
the wave function $\psi$ satisies the system of the discrete Moutard equations
\begin{equation} \label{eq:d-Moutard-ag}
\psi_{(ij)}(P) -  \psi(P) = f^{ij}(\psi_{(i)}(P) - \psi_{(j)}(P) ), 
\qquad 1\leq i < j \leq N,
\end{equation}
where 
\begin{equation} \label{eq:f-ij-alg-def}
f^{ij} = \lim_{P\to Q_i}\frac{\psi_{(ij)}(P)}{\psi_{(i)}(P)} =
-\lim_{P\to Q_j}\frac{\psi_{(ij)}(P)}{\psi_{(j)}(P)} , 
\qquad i < j .
\end{equation}
To obtain B-quadrilateral lattices we pick up $M+1$ points $P_1,\dots
,P_{M+1}\in\Gamma$ of the Riemann surface. Then $x_j(m) = \Psi(m|P_j)$, $1\leq j \leq
M+1$ serve as homogeneous coordinates of the lattice. However in this way we obtain
B-quadrilateral lattices in complex projective space. In order to get real
lattices certain additional restrictions, which were also given in \cite{DGNS}, 
should be imposed on the algebro-geometric data.

In \cite{DGNS} the multidimensional aspects of the system
\eqref{eq:d-Moutard-ag} were not of particular importance. Also the role of the
Prym variety and of the corresponding theta function was not fully exploited.
Our goal here is to fill this point. We start from the
imediate consequence of 
Definition \eqref{def:Prym-variety} of the Prym variety.
\begin{Cor}
Denote by $D(m)$ the divisor of additional zeros of $\psi(m)$, then
\begin{equation}
D(m) - D \stackrel{\mathcal{J}({\hat{\Gamma}})}{=}
\sum_{k=1}^N m_k \left(\sigma(Q_k) - Q_k   \right)
\end{equation}
moves linearly within the Prym variety.
\end{Cor}

An important part of the algebro-geometric theory of integrable systems consists
on providing the explicit formulas, in terms of the Riemann theta functions of
the corresponding Jacobi varieties, for the wave functions and the soliton
fields. In the case of  the special Riemann surfaces used in the paper, there
exist \cite{Fay} formulas connecting the theta functions of $\hat\Gamma$,
$\Gamma$ and $\mathcal{P}_\sigma$. However, instead of reducing
the explicit expressions given in \cite{AKV,DGNS} for the generic curves,  
we will follow the reasoning of \cite{DJKM-Prym-BKP}. 
In order to present the explicit
formulas, in terms of the (Riemann--) Prym theta function, for the wave function
and other relevant data we will use Lema~\ref{lem:R-P}. 

Let us define
\begin{equation*}
\mathbf{V}_k = \int_{Q_\infty}^{Q_k}\!\!\!\boldsymbol{w} \in\mathbb{C}^g, 
\qquad 1\leq k \leq N, \qquad \boldsymbol{w} = (w_1,\dots , w_g),
\end{equation*}
then equations \eqref{eq:w} and \eqref{eq:omST-b} imply that
\begin{equation} \label{eq:V-om}
\phi(\mathbf{V}_k) = \int_{\sigma(Q_k)}^{Q_k}\!\!\! \boldsymbol{\omega} =
-\frac{1}{2\pi i}\oint_{\boldsymbol{b}}\omega_{\sigma(Q_k) Q_k},
\qquad \boldsymbol{b} = (b_1,\dots , b_{2g}).
\end{equation}
\begin{Prop}
The BQL wave function $\psi(m)$ can be written down 
with the help of the Prym theta functions as follows 
\begin{equation} \label{eq:psi}
\psi(m|P)= \frac{\theta ( \int_{Q_\infty}^P\!\! \boldsymbol{w}- 
\sum_{k=1}^N m_k \mathbf{V}_k - \mathbf{e} ;\Pi )\, \theta( \mathbf{e};\Pi )}
{\theta ( \sum_{k=1}^N m_k \mathbf{V}_k + \mathbf{e} ;\Pi) \,
\theta( \int_{Q_\infty}^P\!\! \boldsymbol{w}- \mathbf{e};\Pi )}  
\exp \left( \sum_{k=1}^N m_k \int_{Q_\infty}^P \!\!\!
\omega_{\sigma(Q_k) Q_k} \right), 
\end{equation}
where 
\begin{equation} \label{eq:Z}
\phi(\mathbf{e}) = \mathbf{A}(D) -\mathbf{A}(Q_0) + \pi^*(\mathbf{K}_\Gamma) .
\end{equation}
\end{Prop}
\begin{proof}
Using the property \eqref{eq:D-C} of the divisor $D$, the Hurwitz formula
\begin{equation}
C_{\hat{\Gamma}} \stackrel{\mathcal{J}_{2\hat{g} -2}(\Gamma)}{=}  
\pi^* C_\Gamma + Q_0 + Q_\infty
\end{equation}
relating canonical divisors on $\hat{\Gamma}$ and $\Gamma$, and the relation
\begin{equation}
\mathbf{A}(C_{\hat{\Gamma}}) = - 2 \mathbf{K}_{\hat{\Gamma}}
\end{equation}
between the canonical divisor and the Riemann constants vector, we obtain
\begin{equation*}
\sigma(\mathbf{A}(D) -\mathbf{A}(Q_0) + \pi^*(\mathbf{K}_\Gamma)) 
=-\mathbf{A}(D) +\mathbf{A}(Q_0) - \pi^*(\mathbf{K}_\Gamma),
\end{equation*}
which asserts that the definition of the vector $\mathbf{e}$ in \eqref{eq:Z} 
is meaningful. 

To show that the right hand side of equation \eqref{eq:psi} is single valued on
$\hat{\Gamma}$ we check that it is independent on the integration path in 
the integrals $\int_{Q_\infty}^P$. When two paths differ by an elementary cycle 
we use the properties \eqref{eq:a}-\eqref{eq:Pi} of the holomorphic
differentials,
the quasi-periodicity properties of the theta
functions
\begin{equation*}
\theta(\mathbf{z}+\mathbf{e}_k;{\Pi}) = \theta(\mathbf{z};{\Pi}),\qquad
\theta(\mathbf{z}+\Pi \mathbf{e}_k;{\Pi}) = \exp(-\pi i \Pi_{kk} - 2\pi i z_k)
\theta(\mathbf{z};{\Pi}),
\end{equation*}
where $\mathbf{e}_k$ are vectors of the standard basis in $\mathbb{C}^g$,
and the relations \eqref{eq:omST-a} and \eqref{eq:V-om}. 
From now on our path of integration avoids the cuts, like in fromulas
\eqref{eq:omST-b}-\eqref{eq:omST-PQ}.

As the normalizaton condition at $Q_\infty$ is obvious (the theta function is
even) we are left with the analyticity properties. Lemma \ref{lem:R-P} implies
that the right hand side has simple poles at points of the divisor $D$. Apart
from the zeros of the theta function in the nominator (which may
eventually cancel with the poles at $D$), the only other
poles and zeros are consequences of the analytical properties of the integral in
the exponential part. Let us choose a local parameter $z_k(P)$ at $Q_k$, then 
\begin{equation*}
\omega_{\sigma(Q_k) Q_k}(P) \stackrel{P\to Q_k}{=}\left(-\frac{1}{z_k(P)} +
\dots \right)dz_k(P),
\end{equation*}
which implies that
\begin{equation*}
\int_{Q_\infty}^P \!\!\!
\omega_{\sigma(Q_k) Q_k}  \stackrel{P\to Q_k}{=} - \log z_k(P) + O(1),
\end{equation*}
and, in consequence, the right hand side in equation \eqref{eq:psi} has pole
of order $m_k$ at $Q_k$. Similarly, since
$z_k(\sigma(P))$ is a local parameter at
$\sigma(Q_k)$, we have
\begin{equation*}
\int_{Q_\infty}^P \!\!\!
\omega_{\sigma(Q_k) Q_k}  
\stackrel{P\to \sigma(Q_k)}{=} \log z_k(\sigma(P)) + O(1),
\end{equation*}
and the right hand side in equation \eqref{eq:psi} has zero
of order $m_k$ at $\sigma(Q_k)$.
\end{proof}
\begin{Cor}
The potentials read
\begin{equation} \label{eq:f-ij-alg}
f^{ij}(m) = 
\frac{\theta (\mathbf{V}_i +\sum_{k=1}^N m_k \mathbf{V}_k +\mathbf{e} ;\Pi )
\theta (\mathbf{V}_j +\sum_{k=1}^N m_k \mathbf{V}_k +\mathbf{e} ;\Pi )}
{\theta ( \sum_{k=1}^N m_k \mathbf{V}_k + \mathbf{e} ;\Pi) \,
\theta (\mathbf{V}_i + \mathbf{V}_j + 
\sum_{k=1}^N m_k \mathbf{V}_k +\mathbf{e} ;\Pi )}  \lambda_{ij}^{-1}, \quad i<j
\end{equation}
where
\begin{equation}
\lambda_{ij} = 
\exp \left( \int_{Q_\infty}^{Q_i} \!\!\! \omega_{\sigma(Q_j) Q_j} \right),
\qquad i<j.
\end{equation} 
The BQL (the discrete BKP) $\tau$-function within the above class of
solutions reads
\begin{equation} \label{eq:tau_alg}
\tau(m) = \theta ( \sum_{k=1}^N m_k \mathbf{V}_k + \mathbf{e} ;\Pi)
\prod_{i<j} \lambda_{ij}^{m_i m_j}.
\end{equation}
\end{Cor}
\begin{proof}
Expression \eqref{eq:f-ij-alg} for the potentials is a direct consequence of
their algebro-geometric definition \eqref{eq:f-ij-alg-def} and of equation
\eqref{eq:psi}. Then the formula \eqref{eq:tau_alg}
for $\tau$-function follows easily from its
definiton \eqref{eq:tau}.
\end{proof}
\begin{Rem}
From general considerations of \cite{DGNS} we know that
\begin{equation} \label{eq:lambda}
\lambda_{ij} = 
\exp \left( \int_{Q_\infty}^{Q_i} \!\!\! \omega_{\sigma(Q_j) Q_j} \right) =
-\exp \left( \int_{Q_\infty}^{Q_j} \!\!\! \omega_{\sigma(Q_i) Q_i} \right),
\quad i<j,
\end{equation} 
which reflects the second equality in \eqref{eq:f-ij-alg-def}. 
\end{Rem}
\section{Transformations of the B-quadrilateral lattice}
\label{sec:B-transf}
Below we present the reduction of the vectorial fundamental transformation 
compatible with the B-quadrilateral lattice constraint.
In literature
\cite{NiSchief} there is known the direct vectorial
Moutard transformation between solutions of the  BQL linear problem 
\eqref{eq:BKP-lin} providing thus the corresponding transformation between
solutions
of the discrete BKP equation \eqref{eq:BKP-nlin}. Our goal will be to find the
transition to the Pfaffian expressions of \cite{NiSchief} starting from the BQL
reduction of the fundamental transformation. 
In describing this connection we follow the
ideas of \cite{LiuManas-BKP}, where similar problem between the Grammian
expressions for binary Darboux transformation of the KP hierarchy has
been transformed, in the BKP reduction, into the Pfaffian form
\cite{Hirota-BKP-Pf} (see also \cite{Hirota-BKP-Pf,TsujimotoHirota} for other
aspects of the relation of Pfaffians with 
the BKP hierarchy and the discrete BKP
equation).

\subsection{The fundamental transformation of the QL}
Let us first recall some basic facts concerning the vectorial fundamental
transformation of the quadrilateral lattice.
Geometrically, the (scalar) fundamental transformation is the relation 
between two 
quadrilateral
lattices $x$ and $\hat{x}$ such that for each direction $i$ the
points $x$, $\hat{x}$, $x_{(i)}$ and $\hat{x}_{(i)}$ are coplanar. 

We
present below the algebraic description of its vectorial extension (see
\cite{MDS,TQL,MM} for details) in the affine formalism.
Given the solution $\boldsymbol{Y}_i:\mathbb{Z}^N\to\mathbb{V}$, 
$\mathbb{V}$ being a
linear space, of the linear system \eqref{eq:lin-X}, and given the solution 
$\boldsymbol{Y}^*_i:\mathbb{Z}^N\to\mathbb{V}^*$, $\mathbb{V}^*$ being the 
dual of  $\mathbb{V}$, of the linear system \eqref{eq:lin-h}. These allow to
construct the linear operator valued potential 
$\boldsymbol{\Omega}(\boldsymbol{Y},\boldsymbol{Y}^*):
\mathbb{Z}^N\to L(\mathbb{V})$,
defined by 
\begin{equation} \label{eq:Omega-Y-Y}
\Delta_i \boldsymbol{\Omega}(\boldsymbol{Y},\boldsymbol{Y}^*) = 
\boldsymbol{Y}_i \otimes\boldsymbol{Y}^*_i, 
\qquad i = 1,\dots , N;
\end{equation} 
similarly, one defines 
$\boldsymbol{\Omega}(\boldsymbol{X},\boldsymbol{Y}^*):
\mathbb{Z}^N\to L(\mathbb{V},\mathbb{R}^M)$ and 
$\boldsymbol{\Omega}(\boldsymbol{Y},h):
\mathbb{Z}^N\to \mathbb{V}$ by
\begin{align} \label{eq:Omega-X-Y}
\Delta_i \boldsymbol{\Omega}(\boldsymbol{X},\boldsymbol{Y}^*) & = 
\boldsymbol{X}_i \otimes\boldsymbol{Y}^*_i, \\
\Delta_i\boldsymbol{\Omega}(\boldsymbol{Y},h)  & =
\boldsymbol{Y}_i \otimes h_i. 
\end{align} 
If $\boldsymbol{\Omega}(\boldsymbol{Y},\boldsymbol{Y}^*)$ is invertible then
the (vectorial) fundamental transform of the lattice $\bx$ is given by
\begin{equation} \label{eq:fund-vect}
\hat{\bx}  = \bx - 
\boldsymbol{\Omega}(\boldsymbol{X},\boldsymbol{Y}^*)
\boldsymbol{\Omega}(\boldsymbol{Y},\boldsymbol{Y}^*)^{-1}
\boldsymbol{\Omega}(\boldsymbol{Y},h).
\end{equation}
The corresponding transformation of the $\tau$-function $\tilde\tau$ of the quadrilateral 
lattice reads
\begin{equation} \label{eq:fund-vect-tau}
\hat{\tilde\tau} = \tilde\tau \det
\boldsymbol{\Omega}(\boldsymbol{Y},\boldsymbol{Y}^*).
\end{equation}
When $\dim\mathbb{V}=1$ we obtain the formula relating the quadrilateral
lattice $\bx$ and its fundmental transform $\hat{\bx}$. The vectorial
fundamental transformation can be considered as superposition of
$\dim\mathbb{V}$ (scalar) fundamental transformations; on intermediate stages
the rest of the transformation data should be suitably transformed as well.
Such a description contains already the principle of permutability of such
transformations, which follows from the following observation~\cite{TQL}.
\begin{Lem}
Assume the following splitting of the data of the vectorial fundamental
transformation
\begin{equation}
\boldsymbol{Y}_i = \left( \begin{array}{c} 
\boldsymbol{Y}_i^a \\ \boldsymbol{Y}_i^b \end{array} \right),\qquad
\boldsymbol{Y}_i^* = \left( \begin{array}{cc} 
\boldsymbol{Y}_{ai}^{*}\; , & \boldsymbol{Y}_{b i}^{*} \end{array} \right),
\end{equation}
associated with the partition $\mathbb{V} = \mathbb{V}_a \oplus \mathbb{V}_b$,
which implies the following splitting of the potentials
\begin{equation} \label{eq:split-fund-1}
\boldsymbol{\Omega}(\boldsymbol{Y},h) =  \left( \begin{array}{c} 
\boldsymbol{\Omega}(\boldsymbol{Y}^a,h) \\ 
\boldsymbol{\Omega}(\boldsymbol{Y}^b,h) \end{array} \right), \qquad
\boldsymbol{\Omega}(\boldsymbol{Y},\boldsymbol{Y}^*) = \left( \begin{array}{cc}
\boldsymbol{\Omega}(\boldsymbol{Y}^a,\boldsymbol{Y}_a^*) &
\boldsymbol{\Omega}(\boldsymbol{Y}^a,\boldsymbol{Y}_b^*) \\
\boldsymbol{\Omega}(\boldsymbol{Y}^b,\boldsymbol{Y}_a^*) &
\boldsymbol{\Omega}(\boldsymbol{Y}^b,\boldsymbol{Y}_b^*)\end{array} \right), 
\end{equation}
\begin{equation} \label{eq:split-fund-2}
\boldsymbol{\Omega}(\boldsymbol{X},\boldsymbol{Y}^*) = \left( \begin{array}{cc} 
\boldsymbol{\Omega}(\boldsymbol{X},\boldsymbol{Y}_a^*)\; , &
\boldsymbol{\Omega}(\boldsymbol{X},\boldsymbol{Y}_b^*)\end{array} \right).
\end{equation} 
Then the vectorial fundamental tansformation is equivalent to the following
superposition of vectorial fundamental transformations:\\
1) Transformation $\bx\to\hat{\bx}^{\{a\}}$ with the potentials
$\boldsymbol{\Omega}(\boldsymbol{Y}^a,h)$, 
$\boldsymbol{\Omega}(\boldsymbol{Y}^a,\boldsymbol{Y}_a^*)$, 
$\boldsymbol{\Omega}(\boldsymbol{X},\boldsymbol{Y}_a^*)$
\begin{equation} \label{eq:fund-vect-a}
\hat{\bx}^{\{a\}}  = \bx - 
\boldsymbol{\Omega}(\boldsymbol{X},\boldsymbol{Y}^*_a)
\boldsymbol{\Omega}(\boldsymbol{Y}^a,\boldsymbol{Y}^*_a)^{-1}
\boldsymbol{\Omega}(\boldsymbol{Y}^a,h).
\end{equation}
2) Application on the result the vectorial fundamental transformation with the
transformed potentials
\begin{equation} \label{eq:fund-vect-a-b}
\hat{\bx}^{\{a,b\}}  = \hat{\bx}^{\{a\}} - 
\hat{\boldsymbol{\Omega}}(\boldsymbol{X},\boldsymbol{Y}^*_b)^{\{a\}}
[\hat{\boldsymbol{\Omega}}(\boldsymbol{Y}^b,\boldsymbol{Y}^*_b)^{\{a\}}]^{-1}
\hat{\boldsymbol{\Omega}}(\boldsymbol{Y}^b,h)^{\{a\}},
\end{equation}
where
\begin{align} 
\hat{\boldsymbol{\Omega}}(\boldsymbol{Y}^b,h)^{\{a\}} & =
\boldsymbol{\Omega}(\boldsymbol{Y}^b,h) - 
\boldsymbol{\Omega}(\boldsymbol{Y}^b,\boldsymbol{Y}^*_a)
\boldsymbol{\Omega}(\boldsymbol{Y}^a,\boldsymbol{Y}^*_a)^{-1}
\boldsymbol{\Omega}(\boldsymbol{Y}^a,h),
\\
\hat{\boldsymbol{\Omega}}(\boldsymbol{Y}^b,\boldsymbol{Y}^*_b)^{\{a\}} & =
\boldsymbol{\Omega}(\boldsymbol{Y}^b,\boldsymbol{Y}^*_b) - 
\boldsymbol{\Omega}(\boldsymbol{Y}^b,\boldsymbol{Y}^*_a)
\boldsymbol{\Omega}(\boldsymbol{Y}^a,\boldsymbol{Y}^*_a)^{-1}
\boldsymbol{\Omega}(\boldsymbol{Y}^a,\boldsymbol{Y}^*_b),
\\
\hat{\boldsymbol{\Omega}}(\boldsymbol{X},\boldsymbol{Y}^*_b)^{\{a\}}  & =
\boldsymbol{\Omega}(\boldsymbol{X},\boldsymbol{Y}^*_b) - 
\boldsymbol{\Omega}(\boldsymbol{X},\boldsymbol{Y}^*_a)
\boldsymbol{\Omega}(\boldsymbol{Y}^a,\boldsymbol{Y}^*_a)^{-1}
\boldsymbol{\Omega}(\boldsymbol{Y}^a,\boldsymbol{Y}^*_b).
\label{eq:fund-vect-potentials-slit}
\end{align}
\end{Lem}
\begin{Cor}
The normalized tangent vectors $\boldsymbol{X}_i$ and the Lam\'{e} coefficients
$h_i$ are transformed, at the intermediate step, according
to formulas
\begin{align}
\hat{\boldsymbol{X}}_i^{\{a\}} & = \boldsymbol{X}_i -
\boldsymbol{\Omega}(\boldsymbol{X},\boldsymbol{Y}^*_a)
\boldsymbol{\Omega}(\boldsymbol{Y}^a,\boldsymbol{Y}^*_a)^{-1}
\boldsymbol{Y}^a_i,
\\
\hat{h}_i^{\{a\}} & = h_i - \boldsymbol{Y}^*_{i a}
\boldsymbol{\Omega}(\boldsymbol{Y}^a,\boldsymbol{Y}^*_a)^{-1}
\boldsymbol{\Omega}(\boldsymbol{Y}^a,h),
\end{align}
which also give the corresponding transforms of the second set of
transformation data 
$\boldsymbol{Y}^b$ and $\boldsymbol{Y}^*_b$
\begin{align}
\hat{\boldsymbol{Y}}_i^{b\{a\}} & = \boldsymbol{Y}_i^b -
\boldsymbol{\Omega}(\boldsymbol{Y}^b,\boldsymbol{Y}^*_a)
\boldsymbol{\Omega}(\boldsymbol{Y}^a,\boldsymbol{Y}^*_a)^{-1}
\boldsymbol{Y}^a_i,
\\
\hat{\boldsymbol{Y}}_{i b}^{*\{a\}} & = \boldsymbol{Y}_{i b}^* - 
\boldsymbol{Y}^*_{i a}
\boldsymbol{\Omega}(\boldsymbol{Y}^a,\boldsymbol{Y}^*_a)^{-1}
\boldsymbol{\Omega}(\boldsymbol{Y}^a, \boldsymbol{Y}_{b}^*);
\end{align}
which agree with the transformation rules 
\eqref{eq:fund-vect-potentials-slit} for the potentials, i.e., 
\begin{align*} 
\hat{\boldsymbol{\Omega}}(\boldsymbol{Y}^b,h)^{\{a\}} &=
\boldsymbol{\Omega}(\hat{\boldsymbol{Y}}^{b\{a\}},\hat{h}^{\{a\}}), \\
\hat{\boldsymbol{\Omega}}(\boldsymbol{Y}^b,\boldsymbol{Y}^*_b)^{\{a\}} &=
\boldsymbol{\Omega}(\hat{\boldsymbol{Y}}^{b\{a\}},\hat{\boldsymbol{Y}}_b^{*\{a\}}),
\\
\hat{\boldsymbol{\Omega}}(\boldsymbol{X},\boldsymbol{Y}^*_b)^{\{a\}}  & =
\boldsymbol{\Omega}(\hat{\boldsymbol{X}}^{\{a\}},\hat{\boldsymbol{Y}}_b^{*\{a\}}).
\end{align*}
\end{Cor}
\begin{Rem}
The same result $\hat\bx = \hat{\bx}^{\{a,b\}}=\hat{\bx}^{\{b,a\}}$
is obtained exchanging the order of transformations, exchanging also the indices
$a$ and $b$ in formulas 
\eqref{eq:fund-vect-a}-\eqref{eq:fund-vect-potentials-slit}.
\end{Rem}

\begin{Rem}
If we denote by $\hat{\bx}^{\{1,2\}}$ the
quadrilateral lattice obtained by superposition of two (scalar) fundamental 
transforms from $\bx$ to $\hat{\bx}^{\{1\}}$ and $\hat{\bx}^{\{2\}}$, then the points
$\bx$, $\hat{\bx}^{\{1\}}$, $\hat{\bx}^{\{2\}}$ and $\hat{\bx}^{\{1,2\}}$ are coplanar
again, i.e., the fundamental transformations reproduce the planarity constraint
responsible for integrability of the quadrilateral lattice.
\end{Rem}

\subsection{The BQL (Moutard) reduction of the fundamental transformation}
In this section we describe restrictions on the data of the fundamental
transformation in order to preserve the reduction from QL to BQL. 
As usuall (see, for example \cite{TQL,q-red,MM}) the reduction of the
fundamental transformation for the special quadrilateral lattices 
mimics the geometric properties of the lattice. Because the basic geometric
property of the (scalar) fundamental transformation can be interpreted as
construction of a "new level" of the quadrilateral lattice, then it is natural 
to define the reduced transformation in a similar spirit. Our definition of  
\emph{the BQL reducion of the fundamental transformation} is therefore based on
the following observation. 
\begin{Lem} \label{lem:BQL-fund}
Given B-quadrilateral lattice $x:\ZZ^N\to\PP^M$ and its fundamental transform
$\hat{x}$ constructed under additional assumption that for any point
$x$ of the lattice and any pair $i,j$ of different directions, the four points
$x$, $x_{(ij)}$, $\hat{x}_{(i)}$ and $\hat{x}_{(j)}$ are coplanar. Then the
lattice $\hat{x}:\ZZ^N\to\PP^M$ is B-quadrilateral lattice as well.
\end{Lem}
\begin{proof}
The result is equivalent to the $4$-dimensional consistency of the BQL lattice.
Indeed, in Proposition~\ref{lem:4D-consist-BQL} let the forth direction be
identified withe the transformation direction (the three first directions are
the lattice directions $i$, $j$, $k$). Then the implication
$x_4 \in \langle x_{124}, x_{134}, x_{234} \rangle$ is rewritten in the form
$\hat{x} \in \langle \hat{x}_{(ij)}, \hat{x}_{(ik)}, \hat{x}_{(jk)} \rangle$.
\end{proof}
\begin{Def}
The fundamental transform $\hat{x}$ of a B-quadrilateral lattice $x:\ZZ^N\to\PP^M$  
constructed under additional assumption that for any point
$x$ of the lattice and any pair $i,j$ of different directions, the four points
$x$, $x_{(ij)}$, $\hat{x}_{(i)}$ and $\hat{x}_{(j)}$ are coplanar is called
\emph{the BQL reduction} of the fundamental transformation of $x$.
\end{Def}

On the algebraic level, the Darboux-type transformation of the solutions
of the linear problem \eqref{eq:BKP-lin} was introduced and studied by Nimmo
and Schief in \cite{NiSchief} as \emph{discretization of the Moutard 
transformation}. We will derive their results from the general theory of 
transformations of the quadrilateral lattice.
\begin{Lem} \label{lem:Y-Y*-BQL}
Given a scalar solution $Y_i$ of the linear problem \eqref{eq:lin-X} with the
rotation coefficients restricted by the BQL
reduction \eqref{eq:beta-BQL-ij}-\eqref{eq:beta-BQL-ji}, 
denote by $\theta =  \Omega(Y,h)$ the corresponding potential, where the
Lam\'{e} coefficients are given by equation \eqref{eq:h-tau}. Then
the functions
\begin{equation} \label{eq:Yi*-BQL}
Y_i^* =  (-1)^{\sum_{k<i}m_k}\frac{\tau}{\tau_{(i)}} (\theta + \theta_{(i)})
\end{equation}
are solutions of the adjoint linear problem \eqref{eq:lin-h}
in the BQL reduction, and
the function $\theta^2$ can be taken as the corresponding potential
$\Omega(Y,Y^*)$
\begin{equation} \label{eq:Om-BQL}
\theta^2 = \Omega(Y,Y^*).
\end{equation}
\end{Lem}
\begin{proof}
By direct calculation one verifies that the functions defined in
\eqref{eq:Yi*-BQL} satisfy the reduced system \eqref{eq:lin-h}. Similarly one
checks validity of equation \eqref{eq:Om-BQL}.
\end{proof}
\begin{Rem}
Notice that equation \eqref{eq:Om-BQL} implies, under assumptions of
Lemma~\ref{lem:Y-Y*-BQL}, the form \eqref{eq:Yi*-BQL} of the solution of the
adjoint linear problem.
\end{Rem}
\begin{Prop} \label{prop:BQL-fund}
Given BQL lattice $x$ with homogeneous representation $\bx$ in the gauge of the
linear problem \eqref{eq:BKP-lin}, then the transform of $\bx$ 
constructed using equation~\eqref{eq:fund-vect}
with the data described in Lemma~\ref{lem:Y-Y*-BQL} data satisfies 
the conditions of the BQL reduction.
\end{Prop}
\begin{proof}
The fundamental transform of
$\bx$ constructed with such a data reads
\begin{equation} \label{eq:Omega-XY-BQL}
\hat\bx = \bx - \boldsymbol{\Omega}(\bX,Y^*)/\theta.
\end{equation}
Equation \eqref{eq:Omega-X-Y}, with $\boldsymbol{\Omega}(\bX,Y^*)$ given above, 
can be rewritten then in the following
form 
\begin{equation} \label{eq:Moutard-transf}
\hat\bx_{(i)} - \bx = \frac{\theta}{\theta_{(i)}}\left( \hat\bx -
\bx_{(i)}\right),
\end{equation}
which, together with the linear problem \eqref{eq:BKP-lin}, implies the 
linear relation 
\begin{equation}
\frac{\theta}{\theta_{(i)}}\left(\hat\bx_{(i)} - \bx \right) -
\frac{\theta}{\theta_{(j)}}\left(\hat\bx_{(j)} - \bx \right) +
\frac{1}{f^{ij}}\left(\bx_{(ij)} - \bx \right) =0, \quad i< j,
\end{equation}
between the homogeneous coordinates of the points $x$, $x_{(ij)}$ 
of the lattice and the points $\hat{x}_{(i)}$ and $\hat{x}_{(j)}$
of its fundamental transform, which is the 
algebraic expression of their coplanarity.
\end{proof}
In the approach of \cite{NiSchief} the Moutard transform $\hat\bx$ of $\bx$ was
defined in terms of the system \eqref{eq:Moutard-transf}. Then $\hat\bx$
satisfies new Moutard equations \eqref{eq:BKP-lin} with new potential
\begin{equation}
\hat{f}^{ij} = f^{ij}\frac{\theta_{(i)}\theta_{(j)}}{\theta\theta_{(ij)}},
\qquad i<j,
\end{equation}
and new $\tau$-function
\begin{equation} \label{eq:transf-Mout-tau}
\hat\tau = \theta\tau.
\end{equation}
Another important ingredient of \cite{NiSchief} was the existence of the
potential $S(\theta|\bx) = \theta \hat\bx $ which satisfies
the system 
\begin{equation}
\Delta_{i}S(\theta|\bx) = \theta_{(i)} \bx - \theta \bx_{(i)}.
\end{equation}

We have shown that the algebraic reduction, described in
Lemma~\ref{lem:Y-Y*-BQL}, of the data of the fundamental transformation
can be interpreted as a BQL reduction of the transformation.
We close this Section by showing that the above algebraic description 
holds generally.
\begin{Prop}
Any BQL-reduction of the fundamental transformation can be 
algebraically described as
in Proposition~\ref{prop:BQL-fund}. 
\end{Prop}
\begin{proof}
We will follow the
reasoning of \cite{BobSur} used to the same linear problem \eqref{eq:BKP-lin}
but in different geometric context.
Because the BQL-reduced
fundamental transformation can be considered as construction of the new level of
the B-quadrilateral lattice, its algebraic representation 
should be (in appropriate gauge) in the form of the BQL linear problem
\eqref{eq:BKP-lin}
\begin{equation} \label{eq:BQL-transf-lin}
\hat\bx_{(i)} - \bx = f^{0i}(\hat\bx - \bx_{(i)}),
\end{equation}
where we can label the transformation direction by the index $0$. The
compatibility of system \eqref{eq:BQL-transf-lin} gives 
the following equations (compare with
\eqref{eq:nonl-BQL-f})
\begin{equation}
f^{0i}_{(j)}f^{0j} = f^{0j}_{(i)}f^{0i}, \qquad
1 - f^{0j}_{(i)}(f^{0i} + f^{ij}) = - f^{0i}_{(j)}f^{ij}.
\end{equation}
First of them implies the existence of a potential $\theta$ such that
\begin{equation*}
f^{0i} = \frac{\theta}{\theta_{(i)}},
\end{equation*}
thus equations \eqref{eq:Moutard-transf}. The second equation rewritten in terms
of the potential implies that
$\theta$ satisfies linear problem \eqref{eq:BKP-lin}, i.e. $\theta=\Omega(Y,h)$,
where $h$ given by \eqref{eq:h-tau} and  $Y_i$ is a solution 
of the linear problem 
\eqref{eq:lin-X} with the
rotation coefficients restricted by the BQL
reduction \eqref{eq:beta-BQL-ij}-\eqref{eq:beta-BQL-ji}. By equation
\eqref{eq:Yi*-BQL} we define the corresponding solution of the adjoint linear
problem. Finally, direct calculation with the help of the Moutard transformation
formulas \eqref{eq:Moutard-transf} show that the potential 
\begin{equation*}
\boldsymbol{\Omega}(\bX,Y^*) = \theta (\bx - \hat\bx),
\end{equation*}
(compare with equation \eqref{eq:Omega-XY-BQL}) does satisfy
equation \eqref{eq:Omega-X-Y}, thus $\Omega(Y,Y^*)$ is of the form given in
equation \eqref{eq:Om-BQL}.
\end{proof}

\subsection{The BQL reduction of the vectorial fundamental transformation}
In this Section we propose the restrictions on the data of the vectorial
fundamental transformation, which are compatible with the BQL reduction.
\begin{Prop} \label{prop:vect-fund-red-BQL}
Given solution $\boldsymbol{Y}_i:\ZZ^N\to\mathbb{V}$ of the linear problem
\eqref{eq:lin-X} corresponding to the BQL linear problem \eqref{eq:BKP-lin}
satisfied by the homogeneous coordinates $\bx$ of the BQL lattice
$x:\ZZ^N\to\PP^M$.
Denote by $\boldsymbol{\Theta} = \boldsymbol{\Omega}(\boldsymbol{Y},h)$ the
corresponding potential, which is also new
vectorial solution of the BQL linear problem \eqref{eq:BKP-lin}.\\ 
1) Then
\begin{equation} \label{eq:BQL-Y*}
\boldsymbol{Y}^*_i = (-1)^{\sum_{k<i}m_k}\frac{\tau}{\tau_{(i)}} 
(\boldsymbol{\Theta}^{\mathrm{t}} + \boldsymbol{\Theta}_{(i)}^{\mathrm{t}})
\end{equation}
provides a vectorial
solution of the adjoint linear problem, and the corresponding 
potential
$\boldsymbol{\Omega}(\boldsymbol{Y},\boldsymbol{Y}^*)$ allows for the following
constraint
\begin{equation} \label{eq:vect-Moutard-constr}
\boldsymbol{\Omega}(\boldsymbol{Y},\boldsymbol{Y}^*) + 
\boldsymbol{\Omega}(\boldsymbol{Y},\boldsymbol{Y}^*)^{\mathrm{t}} =
2 \boldsymbol{\Theta}\otimes \boldsymbol{\Theta}^{\mathrm{t}}.
\end{equation}
2) The fundamental vectorial transform $\hat\bx$ of $\bx$, given by
\eqref{eq:fund-vect} with the potentials $\boldsymbol{\Omega}$ restricted as
above can be considered as the superposition of $\dim\mathbb{V}$ (scalar)
discrete BQL reduced fundamental transforms. 
\end{Prop}
\begin{proof}
The point 1) can be checked by direct calculation.
To prove the point 2) notice that when $\dim\mathbb{V}=1$ we obtain the 
BQL reduction of the fundamental transformation in the
setting of Proposition~\ref{prop:BQL-fund}. For $\dim\mathbb{V}>1$
the statement follows from the standard reasoning applied to superposition of
two reduced vectorial fundamental transformations (compare with \cite{TQL,q-red}). 

Assume the splitting 
$\mathbb{V}=\mathbb{V}_a\oplus\mathbb{V}_b$ of the vectorial space $\mathbb{V}$,
and the induced splitting of the basic data $\boldsymbol{Y}_i$ of the
transformation. Then we have also (in shorthand notation, compare equations
\eqref{eq:split-fund-1}-\eqref{eq:split-fund-2})
\begin{equation}
\boldsymbol{\Theta} = \left(   \begin{array}{c}
\boldsymbol{\Theta}^a \\ \boldsymbol{\Theta}^b \end{array} \right),
\qquad
\boldsymbol{\Omega}(\boldsymbol{Y},\boldsymbol{Y}^*) = 
\left(   \begin{array}{cc}
\boldsymbol{\Omega}^a_a &  \boldsymbol{\Omega}^a_b \\
\boldsymbol{\Omega}^b_a &  \boldsymbol{\Omega}^b_b 
\end{array} \right),
\end{equation}
and the constraint \eqref{eq:vect-Moutard-constr} reads
\begin{equation} \label{eq:vect-Moutard-constr-split} 
\left(   \begin{array}{cc}
\boldsymbol{\Omega}^a_a &  \boldsymbol{\Omega}^a_b \\
\boldsymbol{\Omega}^b_a &  \boldsymbol{\Omega}^b_b 
\end{array} \right) 
+
\left(   \begin{array}{cc}
\boldsymbol{\Omega}^{a \mathrm{t}}_a &  \boldsymbol{\Omega}^{b \mathrm{t}}_a \\
\boldsymbol{\Omega}^{a \mathrm{t}}_b &  \boldsymbol{\Omega}^{b \mathrm{t}}_b 
\end{array} \right)  
= 2
\left(   \begin{array}{cc}
\boldsymbol{\Theta}^a \otimes \boldsymbol{\Theta}^{a \mathrm{t}}&  
\boldsymbol{\Theta}^a \otimes \boldsymbol{\Theta}^{b \mathrm{t}} \\
\boldsymbol{\Theta}^b \otimes \boldsymbol{\Theta}^{a \mathrm{t}} &
\boldsymbol{\Theta}^b \otimes \boldsymbol{\Theta}^{b \mathrm{t}}
\end{array} \right).
\end{equation}
By straightforward algebra, using equations 
\eqref{eq:vect-Moutard-constr-split}, one checks that the transformed potentials
(compare equations \eqref{eq:fund-vect-potentials-slit})
\begin{align}
\boldsymbol{\Omega}^{b \{ a\}}_b & = \boldsymbol{\Omega}^b_b - 
\boldsymbol{\Omega}^b_a [\boldsymbol{\Omega}^a_a]^{-1} \boldsymbol{\Omega}^a_b,
\\
\boldsymbol{\Theta}^{b \{ a\}} & = \boldsymbol{\Theta}^b - 
\boldsymbol{\Omega}^b_a [\boldsymbol{\Omega}^a_a]^{-1} \boldsymbol{\Theta}^a,
\end{align}
satisfy the BQL constraint \eqref{eq:vect-Moutard-constr} as well, i.e.,
\begin{equation}
\boldsymbol{\Omega}^{b \{ a\}}_b +
\boldsymbol{\Omega}^{b \{ a\}\, \mathrm{t} }_b =
2 \, \boldsymbol{\Theta}^{b \{ a\}} \otimes
\boldsymbol{\Theta}^{b \{ a\}\, \mathrm{t} },
\end{equation}
which concludes the proof.
\end{proof}
\begin{Rem}
Because the BQL-reduced fundamental transformation can be cosidered as
construction of new levels of the B-quadrilateral lattice, then
if we denote by $\hat{x}^{\{1,2\}}$ the
B-quadrilateral lattice obtained by superposition of two (scalar) such
transforms from $x$ to $\hat{x}^{\{1\}}$ and $\hat{x}^{\{2\}}$, then 
for each direction $i$ of the lattice the points
$x$, $\hat{x}^{\{1\}}_{(i)}$, $\hat{x}^{\{2\}}_{(i)}$ and 
$\hat{x}^{\{1,2\}}$ are coplanar as well as  the points
$x_{(i)}$, $\hat{x}^{\{1\}}$, $\hat{x}^{\{2\}}$ and $\hat{x}^{\{1,2\}}_{(i)}$.
Similarly, if we consider superpositions of three (scalar) transforms of the
B-quadrilateral lattice $x$ then  
the points $x$, $\hat{x}^{\{1,2\}}$, $\hat{x}^{\{1,3\}}$ and 
$\hat{x}^{\{2,3\}}$ are coplanar as well as  the points
$\hat{x}^{\{1\}}$, $\hat{x}^{\{2\}}$, $\hat{x}^{\{3\}}$ and 
$\hat{x}^{\{1,2,3\}}$.
\end{Rem}

\subsection{The Pfaffian form of the transformation} 
\label{sec:fund-Pf}
Finally, we are going to show that the Pfaffian formulas of the
vectorial discrete Moutard transformation obtained in \cite{NiSchief} 
can be derived from
the corresponding formulas of the fundamental transformation subjected to 
the BQL
reduction. 
 
Denote by $S(\boldsymbol{\Theta} | \boldsymbol{\Theta})$ the antisymmetric part of
$\boldsymbol{\Omega}(\boldsymbol{Y},\boldsymbol{Y}^*)$ then equations 
\eqref{eq:Omega-Y-Y}, \eqref{eq:BQL-Y*} and \eqref{eq:vect-Moutard-constr} imply
\begin{equation}
\Delta_i S(\boldsymbol{\Theta} | \boldsymbol{\Theta}) = 
\boldsymbol{\Theta}_{(i)}\otimes \boldsymbol{\Theta}^{\mathrm{t}} -
\boldsymbol{\Theta}\otimes \boldsymbol{\Theta}^{\mathrm{t}}_{(i)}.
\end{equation}
\begin{Lem}
For $\boldsymbol{\Omega}(\boldsymbol{Y},\boldsymbol{Y}^*)$ and 
$S(\boldsymbol{\Theta} |\boldsymbol{\Theta})$  as above we have
\begin{equation} \label{eq:det-Omega}
\det\boldsymbol{\Omega}(\boldsymbol{Y},\boldsymbol{Y}^*) = 
\det S(\boldsymbol{\Theta} | \boldsymbol{\Theta}) + \left| \begin{array}{cc}
0 & - \boldsymbol{\Theta}^{\mathrm{t}} \\
\boldsymbol{\Theta} &  S(\boldsymbol{\Theta} | \boldsymbol{\Theta})
\end{array} \right|.
\end{equation}
\end{Lem}
\begin{proof}
Notice that the $j$th column $\Omega_j$ of
$\boldsymbol{\Omega}(\boldsymbol{Y},\boldsymbol{Y}^*)$ is of the form
\begin{equation} \label{eq:Omega-j}
\Omega_j = \theta^j \boldsymbol{\Theta} + S_j,
\end{equation}
where $\theta^j$ is the $j$th component of $\boldsymbol{\Theta}$, and
$S_j$ is the $j$th column of $S(\boldsymbol{\Theta} | \boldsymbol{\Theta})$.
Then the basic properties of determinants imply that
\begin{equation} \label{eq:det-Omega-2}
\det \boldsymbol{\Omega}(\boldsymbol{Y},\boldsymbol{Y}^*) =
\det  S(\boldsymbol{\Theta} | \boldsymbol{\Theta}) + \sum_{j=1}^{\dim\mathbb{V}}
\theta^j S(j),
\end{equation}
where by $S(j)$ we denote the matrix  $S(\boldsymbol{\Theta} |
\boldsymbol{\Theta})$ with $j$th column replaced by  $\boldsymbol{\Theta}$. The
second summand in \eqref{eq:det-Omega-2} is the Laplace expansion of that in
\eqref{eq:det-Omega}.
\end{proof}

The standard  properties of determinants of antisymmetric matrices (see
Appendix~\ref{app:Pf}) imply the following result derived in \cite{NiSchief}
directly on the level of vectorial Moutard transformation.
\begin{Cor}
The transformation formula \eqref{eq:fund-vect-tau}  of the QL $\tau$-function
and the relation \eqref{eq:tau-QL-BQL} between both $\tau$-functions imply
the following transformation formula for the BQL $\tau$-function 
\begin{equation}
\hat\tau = \begin{cases}
\tau \: \mathrm{Pf} \: S(\boldsymbol{\Theta} | \boldsymbol{\Theta}) , & 
\dim\mathbb{V} \; \text{even} \\
\tau \: \mathrm{Pf} \left( \begin{array}{cc}
0 & - \boldsymbol{\Theta}^{\mathrm{t}} , \\
\boldsymbol{\Theta} &  S(\boldsymbol{\Theta} | \boldsymbol{\Theta})
\end{array} \right), \qquad & \dim\mathbb{V} \; \text{odd}.
\end{cases}
\end{equation}
\end{Cor}
\begin{Rem}
Notice \cite{NiSchief} that 
\begin{equation}
\left( \begin{array}{cc}
0 & - \boldsymbol{\Theta}^{\mathrm{t}} , \\
\boldsymbol{\Theta} &  S(\boldsymbol{\Theta} | \boldsymbol{\Theta})
\end{array} \right) = S(\boldsymbol{\tilde\Theta} | \boldsymbol{\tilde\Theta}),
\qquad \text{where} \qquad \boldsymbol{\tilde\Theta} = \left( \begin{array}{c}
1 \\ \boldsymbol{\Theta} \end{array} \right),
\end{equation}
which allows to define
\begin{equation}
\mathcal{P}(\boldsymbol{\Theta} ) = \begin{cases}
\mathrm{Pf} \: S(\boldsymbol{\Theta} | \boldsymbol{\Theta}), & 
\dim\mathbb{V} \; \text{even} ,\\
\mathrm{Pf}\: S(\boldsymbol{\tilde\Theta} | \boldsymbol{\tilde\Theta}),
\qquad & \dim\mathbb{V} \; \text{odd},
\end{cases}
\end{equation}
and gives 
\begin{equation}
\hat\tau = \tau \:\mathcal{P}(\boldsymbol{\Theta} ) .
\end{equation}
\end{Rem}

Finally, we will connect the formula of the vectorial fundamental
transformation \eqref{eq:fund-vect} in the BQL reduction
with the Pffafian form of the vectorial Moutard transformation \cite{NiSchief}.
\begin{Cor}
The homogeneous coordinates (in the gauge of the linear problem
\eqref{eq:BKP-lin}) of the BQL lattice $\hat{x}$ obtained from the BQL lattice
$x$ via vectorial transform with the solution $\boldsymbol{\Theta}$ of the
linear problem \eqref{eq:BKP-lin} are given by
\begin{equation}
\hat{x}^i = 
\frac{\mathcal{P}(\boldsymbol{\Theta},x^i)}{\mathcal{P}(\boldsymbol{\Theta})},
\end{equation}
where   
\begin{equation}
\mathcal{P}(\boldsymbol{\Theta},x^i) =
\mathcal{P}\left(\left(\begin{array}{c}\boldsymbol{\Theta}\\x^i \end{array}
\right)\right).
\end{equation}
\end{Cor}
\begin{proof}
We will work using assumptions and notation of
Proposition~\ref{prop:vect-fund-red-BQL}.
Let us define
\begin{equation}
S(\bx | \boldsymbol{\Theta}) = 
\boldsymbol{\Omega}(\boldsymbol{X},\boldsymbol{Y}^*) -
\bx\otimes \boldsymbol{\Theta}^{\mathrm{t}} ,
\end{equation}
then, due to equations \eqref{eq:x-hX}, \eqref{eq:Omega-X-Y} and
\eqref{eq:BQL-Y*} we have
\begin{equation}
\Delta_i S(\bx | \boldsymbol{\Theta}) = 
\bx_{(i)}\otimes \boldsymbol{\Theta}^{\mathrm{t}} -
\bx\otimes \boldsymbol{\Theta}^{\mathrm{t}}_{(i)}.
\end{equation}
By the Cramer rule and equation \eqref{eq:Omega-j}, formula \eqref{eq:fund-vect}
in the considered reduction case can be brought to the form
\begin{equation} \label{eq:Mout-vect}
\hat\bx = \bx - \left( \bx\otimes \boldsymbol{\Theta}^{\mathrm{t}} + 
S(\bx | \boldsymbol{\Theta}) \right)\frac{1}{\det
\boldsymbol{\Omega}(\boldsymbol{Y},\boldsymbol{Y}^*) }\left( \begin{array}{c}
\det S(1) \\ \vdots \\ \det S(\dim \mathbb{V}) \end{array} \right), 
\end{equation}
moreover we have
\begin{equation} \label{eq:theta-Sj}
\boldsymbol{\Theta}^{\mathrm{t}} \left( \begin{array}{c}
\det S(1) \\ \vdots \\ \det S(\dim \mathbb{V}) \end{array} \right) = 
\left| \begin{array}{cc}
0 & - \boldsymbol{\Theta}^{\mathrm{t}} \\
\boldsymbol{\Theta} &  S(\boldsymbol{\Theta} | \boldsymbol{\Theta})
\end{array} \right|.
\end{equation}

Our further analysis splits in the cases of $\dim\mathbb{V}$ being even or odd. 
In the first case the right hand side of
equation \eqref{eq:theta-Sj} vanishes giving
\begin{equation}
\hat\bx = \bx - 
\frac{S(\bx | \boldsymbol{\Theta})}
{\mathrm{Pf} S(\boldsymbol{\Theta}|\boldsymbol{\Theta}) } 
\left( \begin{array}{c}
\mathrm{Pf} S[1] \\ \vdots \\ \mathrm{Pf} S [\dim\mathbb{V}] \end{array} \right),
\end{equation} 
where we used equation \eqref{eq:det-Omega} and the Pfaffian analogue
\eqref{eq:Pfaff-Cramer} of the Cramer rule for 
for solutions of the equation 
$S(\boldsymbol{\Theta}|\boldsymbol{\Theta}) \boldsymbol{y} =
\boldsymbol{\Theta}$. Then the expansion rule for Pfaffians 
\eqref{eq:Laplace-Pfaff} implies that the
$i$th coordinates of the B-quadrilateral lattice $\bx$ and its transform
$\hat\bx$ can be put in the form 
\begin{equation}
\hat{x}^i = \frac{1}{\mathrm{Pf} S(\boldsymbol{\Theta}|\boldsymbol{\Theta})}
\mathrm{Pf}\left(  \begin{array}{ccc}
0 & x^i & S(x^i| \boldsymbol{\Theta} ) \\
-x^i & 0 & -\boldsymbol{\Theta}^{\mathrm{t}} \\
-S(x^i | \boldsymbol{\Theta} )^{\mathrm{t}} & \boldsymbol{\Theta} &
S(\boldsymbol{\Theta} | \boldsymbol{\Theta} ) 
\end{array}  \right) =
\frac{\mathcal{P}(\boldsymbol{\Theta},x^i)}{\mathcal{P}(\boldsymbol{\Theta})}.
\end{equation}

For $\dim\mathbb{V}$ odd, by equations \eqref{eq:det-Omega} and 
\eqref{eq:theta-Sj}, equation \eqref{eq:Mout-vect} reduces to
\begin{equation}
\hat{\boldsymbol{x}} = - \left| \begin{array}{cc}
0 & - \boldsymbol{\Theta}^{\mathrm{t}} \\
\boldsymbol{\Theta} &  S(\boldsymbol{\Theta} | \boldsymbol{\Theta})
\end{array} \right|^{-1} S(\bx|\boldsymbol{\Theta})  
\left( \begin{array}{c}
\det S(1) \\ \vdots \\ \det S(\dim \mathbb{V}) \end{array} \right).
\end{equation}
Expanding $\det S(j)$ with respect to its $j$th column, and using Pfaffian
expressions \eqref{eq:minors-odd-Pf} for the minors of  
$S(\boldsymbol{\Theta} | \boldsymbol{\Theta} ) $ we obtain 
\begin{equation}
\hat{x}^i = \frac{\mathrm{Pf}\left(  \begin{array}{cc}
0 &  S(x^i| \boldsymbol{\Theta} ) \\
-S(x^i | \boldsymbol{\Theta} )^{\mathrm{t}} & 
S(\boldsymbol{\Theta} | \boldsymbol{\Theta} ) 
\end{array}  \right) }
{\mathrm{Pf}\left( \begin{array}{cc}
0 &  -\boldsymbol{\Theta}^{\mathrm{t}}  \\
\boldsymbol{\Theta} & S(\boldsymbol{\Theta} | \boldsymbol{\Theta} ) 
\end{array}  \right) } =
\frac{\mathcal{P}(\boldsymbol{\Theta},x^i)}{\mathcal{P}(\boldsymbol{\Theta})},
\end{equation}
which concludes the proof.
\end{proof}

\section{Conclusion and remarks}
We presented new geometric interpretation of the discrete BKP equation within
the theory of quadrilateral lattices. This new integrable
lattice should be considered, together with the
symmetric lattice \cite{DS-sym} and the quadrilateral 
lattices subject to quadratic
constraints \cite{q-red}, as one of basic reductions of the quadrilateral
lattice. In the forthcomong paper \cite{AD-isoth} we show, for example, that the
discrete isothermic surfaces \cite{BP2} are lattices subjected simultaneously to
the BQL and quadratic (in this case the quadric is the M\"{o}bius sphere)
reductions. 

As in the case of the Hirota (the discrete KP) equation, also the Miwa (the 
discrete BKP)
equation can be considered in the finite fields (or the finite geometry)
setting. In paricular, the main algebro-geometric way of reasoning (see 
\cite{DBK,BD} for the former
discrete KP case) leading to the Prym varieties should be also transferable 
for fields of
characteristic different from two (see \cite{Mumford} for general theory of 
Prym varieties).

The explicit  Prym-theta functional formalae  (it is enough to consider the case
$N=2$) for the wave function and potentials of
the discrete Moutard equation can be also used to provide
characterization of the Prym varieties among all principally polarized abelian 
varieties (the 
Prym--Schottky problem) in the spirit of \cite{Krichever-Prym}. 

\section*{Acknowledgements}
The author would like to thank Jaros{\l}aw Kosiorek and Andrzej
Matra\'{s} for discussions on projective geometry. 

The main part of the paper was
prepared during author's work at DFG Research Center MATHEON in 
Institut f\"{u}r Mathematik of the Technische Universit\"{a}t Berlin.
The paper was supportet also in part by the Polish Ministry of  
Science and Higher Education research grant 1~P03B~017~28. 

\appendix

\section{An alternative proof of the existence of the BQL gauge}

The planarity condition of elementary 
quadrilaterals of QL
can be expressed in terms of generic homogoneous representation 
as the following system of discrete Laplace equations  \cite{MQL}
\begin{equation} \label{eq:Laplace}
\bx_{(ij)} = a^{ij}\bx_{(i)} - a^{ji}\bx_{(j)} + c^{ij}\bx, 
\quad 1\leq i < j \leq N
\end{equation}
whose compatibility are equations
\begin{align} \label{eq:MQL1}
&a^{ij}_{(k)}c^{ik} - a^{ji}_{(k)}c^{jk} =
a^{ik}_{(j)}c^{ij} - a^{ki}_{(j)}c^{jk} = 
a^{jk}_{(i)}c^{ij} - a^{kj}_{(i)}c^{ik}, \\
\label{eq:MQL2}
&a^{ij}_{(k)}a^{ik} = a^{ik}_{(j)}a^{ij} = 
c^{jk}_{(i)} + a^{jk}_{(i)}a^{ij} - a^{kj}_{(i)}a^{ik} \\
\label{eq:MQL3}
&a^{ji}_{(k)}a^{jk} = a^{jk}_{(i)}a^{ji} = 
- c^{ik}_{(j)} + a^{ik}_{(j)}a^{ji} + a^{ki}_{(j)} a^{jk} , \\
\label{eq:MQL4}
&a^{ki}_{(j)}a^{kj} = a^{kj}_{(i)}a^{ki} =
c^{ij}_{(k)} +  a^{ji}_{(k)}a^{kj}
- a^{ij}_{(k)}a^{ki} , 
\end{align}
where $1\leq i < j < k \leq N$. 
Because
\begin{equation}
\bx\wedge\bx_{(ij)}\wedge\bx_{(ik)}\wedge\bx_{(jk)} = 
\bx\wedge\bx_{(i)}\wedge\bx_{(j)}\wedge\bx_{(k)}
(a^{ij}a^{jk}a^{ki} - a^{ji}a^{kj}a^{ik}),
\end{equation}
then the BQL reduction condition is equivalent to
\begin{equation} \label{eq:BKP-cond}
a^{ij}a^{jk}a^{ki} - a^{ji}a^{kj}a^{ik} =0 .
\end{equation} 

We will show that equation \eqref{eq:BKP-cond}
implies existence the gauge function $\rho:\ZZ^N\to\RR$ such that
\begin{align} \label{eq:rho-1}
a^{ij}\rho_{(i)} = a^{ji}\rho_{(j)}, \quad 
a^{ik}\rho_{(i)} & = a^{ki}\rho_{(k)}, \quad
a^{jk}\rho_{(j)} = a^{kj}\rho_{(k)}, \\
\label{eq:rho-2}
\rho_{(ij)} = c^{ij}\rho, \quad
\rho_{(ik)} & = c^{ik}\rho, \quad
\rho_{(jk)} = c^{jk}\rho.
\end{align}
Then, after rescaling $\bx \to \bx/\rho$, the new homogeneous coordinates
satisfy the system \eqref{eq:BKP-lin}.

Let us consider equations \eqref{eq:rho-1}-\eqref{eq:rho-2} as a difference
system, which allows to calculate from $\rho$ and (say) $\rho_{(i)}$ values of
the gauge function in remainig vertices of the hexahedron. 
Notice first that the condition \eqref{eq:BKP-cond} and the system
\eqref{eq:MQL1}-\eqref{eq:MQL4} imply
(it can be verified directly, but actually it follows from Corollary
\ref{cor:BKP-impl}) that
\begin{equation} \label{eq:BKP-impl}
a^{ij}_{(k)}c^{ik}  = a^{ji}_{(k)}c^{jk} , \quad
a^{ik}_{(j)}c^{ij}  = a^{ki}_{(j)}c^{jk} , \quad
a^{jk}_{(i)}c^{ij}  = a^{kj}_{(i)}c^{ik}. 
\end{equation}

The condition \eqref{eq:BKP-cond} assures self-consistency of equations 
\eqref{eq:rho-1}. Then equations \eqref{eq:BKP-impl} imply consistency of
equations \eqref{eq:rho-2} with the following consequence of equations
\eqref{eq:rho-1}
\begin{equation*}
a^{ij}_{(k)}\rho_{(ik)} = a^{ji}_{(k)}\rho_{(jk)}, \quad 
a^{ik}_{(j)}\rho_{(ij)} = a^{ki}_{(j)}\rho_{(jk)}, \quad
a^{jk}_{(i)}\rho_{(ij)} = a^{kj}_{(i)}\rho_{(ik)}. \\
\end{equation*}
Finally, the self-consistency of equations \eqref{eq:rho-2} in finding 
$\rho_{(ijk)}$ follows from equations \eqref{eq:rho-1} and the system 
\eqref{eq:MQL2}-\eqref{eq:MQL4}.

\begin{Rem}
Notice that the system 
\eqref{eq:MQL1}-\eqref{eq:MQL4} of \emph{nonlinear} equations can be 
considered as a system of eight \emph{linear} equations allowing for 
transition 
\begin{equation*}
(a^{ij}, a^{ji}, a^{ik},  a^{ki}, a^{jk},
a^{kj}, c^{ij}, c^{ik},  c^{jk}) \to
(a^{ij}_{(k)}, a^{ji}_{(k)}, a^{ik}_{(j)}, a^{ki}_{(j)},
a^{jk}_{(i)}, a^{kj}_{(i)}, c^{ij}_{(k)}, c^{ik}_{(j)}, c^{jk}_{(i)});
\end{equation*} 
the difference in
the number of unknowns and equations reflects the homogeneous nature of
the linear system \eqref{eq:Laplace}.
\end{Rem}

\section{Pfaffians} \label{app:Pf}
We recall basic properties of Pfaffians \cite{Muir,Proskuryakov}, which we use in 
Section~\ref{sec:fund-Pf}.
Let $A=(a_{ij})_{1\leq i,j \leq 2r}$ be a skew symmetric matrix (i.e.,
$a_{ji} = - a_{ij}$) of the even order $2r$. Consider the form
\begin{equation}
\omega = \sum_{i<j}a_{ij}e_i\wedge e_j,
\end{equation}
then the Pfaffian $\mathrm{Pf}(A)$ of $A$ is defined by
\begin{equation}
\omega^{\wedge r}=
(r!)\;\mathrm{Pf}(A) \; e_1\wedge \dots \wedge e_{2r}.
\end{equation}
For each permutation $\pi$ of $\{1,\dots,2r\}$, put $A^\pi=(a_{\pi(i)\pi(j)})$,
then
\begin{equation}
\mathrm{Pf}(A^\pi) = \mathrm{sgn} \; \pi \; \mathrm{Pf}(A).
\end{equation}
Notice the analogy with the determinant $\det B$ of an arbitrary square matrix
$B=(b_{ij})_{1\leq i,j \leq n}$ expressed in terms of the forms
\begin{equation}
\omega_i = \sum_j b_{ij}e_j 
\end{equation}
as follows
\begin{equation}
\omega_1\wedge \dots \wedge \omega_n = \det (B) \; e_1\wedge \dots \wedge e_{n}.
\end{equation}
It turns out that the determinant of any skew symmetric matrix of an even order
equals to the square of its Pfaffian
\begin{equation}
\det (A) = (\mathrm{Pf}(A))^2.
\end{equation}

For any two subsets $I,J\subset\{1,\dots,n \}$ denote by $B(I,J)$ 
the sub-matrix of $B$
obtained by removing all the $i$th$\in I$ rows and all the
$j$th$\in J$ columns of $B$. 
Then in analogy to the Laplace expansion of determinants
\begin{equation}
\delta_{ij} \det (B) = \sum_{k=1}^n b_{kj}(-1)^{k+i}\det (B(\{k\},\{i\})).
\end{equation} 
we have the following
expansion formula for Pfaffians
\begin{equation} \label{eq:Laplace-Pfaff}
\delta_{ij}\mathrm{Pf}(A) = \sum_{k=1}^{2r} a_{kj}(-1)^{k+i-1}
\mathrm{Pf}(A(\{k,i\},\{k,i\})).
\end{equation}
Both formulas imply 
\begin{equation}
\det (A(\{i\},\{j\})) = - \mathrm{Pf}(A)\; \mathrm{Pf}(A(\{i,j\},\{i,j\})),
\end{equation}
which leads to the following Pfaffian-Cramer rule 
for solutions of the linear
system 
\begin{equation}
A\boldsymbol{y} = \boldsymbol{b}
\end{equation}
with non-degenerate skew symmetrix matrix od the even order:
\begin{equation} \label{eq:Pfaff-Cramer}
y^j = \frac{\mathrm{Pf} (A[j])}{\mathrm{Pf} (A)},
\end{equation}
where by $A[j]$ is denoted the matrix $A$ whose $j$th column is replaced by
\begin{equation}
\boldsymbol{b^\prime} = \left(  b^1,\dots , b^{j-1},
0 , - b^{j+1} , \dots , -b^{2r} \right)^{\mathrm{t}},
\end{equation}
and whose $j$th row is replaced by $-\boldsymbol{b^\prime}^{\mathrm{t}}$.

When the order of the skew symmetric matrix $A$ is odd we have $\det (A) =0$,
but the following formula holds
\begin{equation} \label{eq:minors-odd-Pf}
\det (A(\{i\},\{j\})) = \mathrm{Pf}(A(\{i\},\{i\})) \;\mathrm{Pf}(A(\{j\},\{j\})).
\end{equation}
\bibliographystyle{amsplain}

\begin{thebibliography}{10}

\bibitem{AKV}
A. A. Akhmetishin, I. M. Krichever and Y. S. Volvovski, \emph{Discrete analogues
of the Darboux--Egoroff metrics}, Proc. Steklov Inst. Math. \textbf{225} 16-39.

\bibitem{BD}
M.~Bia{\l}ecki, A.~Doliwa, \emph{Algebro-geometric solution of the discrete 
KP equation over a finite field out of a hyperelliptic curve},  
Comm. Math. Phys.  \textbf{253}  (2005), 157--170.


\bibitem{BP2}
A. I. Bobenko and U. Pinkall, \emph{Discrete isothermic surfaces}, J. Reine Angew.
Math. \textbf{475} (1996) 187--208.

\bibitem{BobSur}
A.~I. Bobenko and Yu.~Suris, \emph{Discrete differential geometry. Consistency 
as integrability}, {\tt math.DG/0504358}.

\bibitem{BBEIM}
E. D. Belokolos, A. I. Bobenko, V. Z. Enol'skii, A. R. Its and V. B. Matveev,
\emph{Algebro-geometric approach to nonlinear integrable equations}, Sronger,
Berlin, 1994.

\bibitem{BoKo}
L. V. Bogdanov and B. G. Konopelchenko, \emph{Lattice and $q$-difference
Darboux--Zakharov--Manakov systems via $\bar\partial$ method}, J. Phys. A: Math.
Gen. \textbf{28} L173--L178.

\bibitem{BoKo-N-KP}
L. V. Bogdanov and B. G. Konoelchenko, \emph{Analytic-bilinear approach to
integrable hiererchies II. Multicomponent KP and 2D Toda hiererchies}, 
J. Math. Phys. \textbf{39} (1998) 4701--4728.

\bibitem{CarrollSpeyer}
G. D. Carroll and D. Speyer, \emph{The cube recurrence},
Electron. J. Combin. \textbf{11} (2004), no. 1, Research Paper 73, 31 pp. 
(electronic).

\bibitem{CDS}
J. Cie\'{s}li\'{n}ski, A. Doliwa and P. M. Santini,
{\it  The Integrable Discrete Analogues of Orthogonal Coordinate Systems
are Multidimensional Circular Lattices},
Phys. Lett. A {\bf 235} (1997) 480--488.


\bibitem{Coxeter}
H.~S.~M.~Coxeter, \emph{Projective Geometry}, Springer, Berlin, 1974.

\bibitem{Darboux-OS}
G. Darboux, \emph{Le\c{c}ons sur les syst\'{e}mes orthogonaux et les
coordonn\'{e}es curvilignes}, Gauthier-Villars, Paris, 1910.

\bibitem{DKJM}
E.~Date, M.~Kashiwara, M. Jimbo and T.~Miwa, \emph{Transformation groups for
soliton equations}, [in:] Proceedings of RIMS Symposioum on Non-Linear
Integrable Systems --- Classical Theory and Quantum Theory (M. Jimbo and T.
Miwa, eds.) World Science Publishing Co., Singapore, 1983, pp. 39--119.


\bibitem{DJKM-BKP}
E.~Date, M. Jimbo, M.~Kashiwara and T.~Miwa, \emph{A new hierarchy of soliton
equations of KP-type},
Physica \textbf{4} D (1982) 343--365.

\bibitem{DJKM-Prym-BKP}
E.~Date, M.~Jimbo, M.~Kashiwara and T.~Miwa, \emph{Quasi-periodic solutions of
the orthognal KP equation -- Transformation groups for soliton equations V},
Publ. RIMS, Kyoto Univ. \textbf{18} (1982) 1111--1119.

\bibitem{q-red}
 A.~Doliwa, \emph{Quadratic reductions of quadrilateral lattices}, J. Geom. 
Phys.
  \textbf{30} (1999), 169--186.

\bibitem{AD-isoth}
A. Doliwa, \emph{Generalized isothermic lattice}, in preparation.

\bibitem{MQL}
A.~Doliwa and P.~M. Santini, \emph{Multidimensional quadrilateral lattices 
are  integrable}, Phys. Lett. A \textbf{233} (1997), 365--372.

\bibitem{DS-sym}
A.~Doliwa and P.~M. Santini, \emph{The symmetric, {D}-invariant and {E}gorov 
reductions of the
  quadrilateral lattice}, J. Geom. Phys. \textbf{36} (2000) 60--102.
 
\bibitem{DS-EMP}
A.~Doliwa and P.~M.~Santini, \emph{Integrable systems and discrete 
geometry}, [in:] Encyclopedia of Mathematical Physics, J. P. Fran\c{c}ois, 
G. Naber and T. S. Tsun (eds.), Elsevier, 2006, Vol. III, pp. 78-87.

\bibitem{DBK}
A.~Doliwa, M.~Bia{\l}ecki, P.~Klimczewski, \emph{The Hirota equation over finite 
fields: algebro-geometric approach and multisoliton solutions}  J. Phys. A
\textbf{36}  (2003) 4827--4839. 

\bibitem{TQL}
A. Doliwa, P. M. Santini and M. Ma{\~n}as,
\emph{Transformations of Quadrilateral Lattices}, J. Math. Phys. \textbf{41}
(2000)  944--990.


\bibitem{DGNS}
A.~Doliwa, P.~Grinevich, M.~Nieszporski, and P.~M. Santini, \emph{Integrable 
lattices and their sub-lattices:
from the discrete Moutard (discrete Cauchy--Riemann) 4-point equation to the 
self-adjoint 5-point scheme}, 
{\tt nlin.SI/0410046}.

\bibitem{Fay}
J.~D.~Fay, \emph{Theta functions on Riemann surfaces}, Springer, Berlin, 1973.

\bibitem{FarkasKra}
H. M. Farkas and I. Kra, \emph{Riemann surfaces}, Springer, Berlin -- New York, 1992.

\bibitem{FominZelevinsky}
S. Fomin and A. Zelevinsky, \emph{The Laurent phenomenon}, Adv. Appl. Math.
\textbf{28} (2002) 119--144.

\bibitem{Hirota-BKP-Pf}
R. Hirota, \emph{Soliton solutions to the BKP equations. I. The Pfaffian
technique}, J. Phys. Soc. Japan \textbf{58} (1989) 2285--2296.

\bibitem{KvL}
V. G. Kac and J. van de Leur, \emph{The $n$-component KP hiererchy and
representation theory}, [in:] Important developments in soliton theory, (A. S.
Fokas and V. E. Zakharov, eds.) Springer, Berlin, 1993, pp. 302--343. 

\bibitem{KoSch-trap}
B.~G. Konopelchenko and W.~K. Schief, \emph{Trapezoidal discrete surfaces: 
geometry and integrability},  J. Geom. Phys.  \textbf{31}  (1999) 75--95.

\bibitem{KoSchiefSBKP}
B.~G. Konopelchenko and W.~K. Schief, \emph{Reciprocal figures, graphical
  statics and inversive geometry of the Schwarzian BKP hierarchy},  
  Stud. Appl. Math.  \textbf{109}  (2002) 89--124.
  
\bibitem{Krichever-Prym}
I.~Krichever, \emph{A characterization of Prym varieties},
\texttt{math.AG/0506238}.  

\bibitem{LiuManas-BKP}
Q. P. Liu and M. Ma\~{n}as, \emph{Pfaffian form of Grammian determinant
solutions of the BKP hierarchy}, {\tt solv-int/9806004}, Chinese Ann. Math. Ser.
A \textbf{23} (2002) 693--698.
 
\bibitem{MM}
M. Ma\~{n}as, \emph{Fundamental transformation for quadrilateral lattices:
first potentials and $\tau$-functions, symmetric and pseudo-Egorov reductions},
J.~Phys. A \textbf{34} (2001) 10413--10421.
 
\bibitem{MDS}
M. Ma\~{n}as, A. Doliwa and P.M. Santini, \emph{Darboux transformations for 
multidimensional quadrilateral lattices. I}, Phys. Lett. A \textbf{232} 
(1997) 99--105.


\bibitem{Miwa} T.~Miwa, \emph{On Hirota's difference equations}, 
  Proc. Japan Acad. \textbf{58} (1982) 9--12.

\bibitem{Moebius}
F. A. M|"{o}bius, \emph{Kann von zwei dreiseitigen Pyramiden eine jede in 
Bezug auf die andere um- und eingeschrieben zugleich heissen?}, 
J. reine angew. Math. \textbf{3} (1828) 273-278. 

\bibitem{Muir}
T. Muir, \emph{A treatise on the theory of determinants}, revised and enlarged
by W. H. Metzler, Dover Publ., New York, 1960.

  
\bibitem{Mumford} 
D.~Mumford, \emph{Prym varieties. I}, Contributions to analysis (a collection of
papers dedicated to Lipman Bers), pp. 325--350, Academic Press, New York, 1974. 

  
\bibitem{NiSchief}
J.~J.~C. Nimmo and W.~K. Schief, \emph{Superposition principles associated with
  the {Moutard} transformation. {An} integrable discretisation of a
  (2+1)-dimensional sine-{Gordon} system}, Proc. R. Soc. London A \textbf{453}
  (1997), 255--279.

\bibitem{Pedoe}
D. Pedoe, \emph{Geometry, a comprehensive course}, Dover Publications, New York,
1988.

\bibitem{Propp}
J. Propp, \emph{The many faces of alternating-sign matrices}, Discrete
Mathematics and 
Theoretical  Computer Science Proceedings \textbf{AA (DM-CCG)} (2001) 43--58.

\bibitem{Proskuryakov}
I. V. Proskuryakov, \emph{Problems in linear algebra}, Mir Publishers, Moscow,
1978.


\bibitem{Sauer}
R.~Sauer, \emph{Differenzengeometrie}, Springer, Berlin, 1970.

\bibitem{Schief-JNMP}
W.~K. Schief, \emph{Lattice geometry of the discrete Darboux, KP, BKP and CKP
equations. Menelaus' and Carnot's theorems}, J. Nonl. Math. Phys. \textbf{10} 
Supplement 2 (2003) 194--208.

\bibitem{Shiota-Prym}
T.~Shiota, \emph{Prym varieties and soliton equations}, 
Infinite dimensional Lie algebras and groups, (V.~G.~Kac, ed.), World
Scientific, Singapore, 1989, pp. 407--448.

\bibitem{Taimanov}
I.~A.~Taimanov, \emph{Prym varieties of branch covers and nonlinear
equations}, Matem. Sbornik \textbf{181} (1990) 934--950.


\bibitem{TsujimotoHirota}
S. Tsujimoto and R. Hirota, \emph{Pfaffian representation of solutions to the
discrete BKP hierarchy in bilinear form}, J. Phys. Soc. Japan \textbf{66} (1997) 
2797--2806.

\bibitem{VeselovNovikov}
A.~P.~Veselov and S.~P.~Novikov, \emph{Finite-gap two-dimensional potential
Schr\"{o}dinger operators: explicit formulae and evolution equations}, Doklady
AN USSR \textbf{279} (1984) 20--24.

\end{thebibliography}

\providecommand{\bysame}{\leavevmode\hbox to3em{\hrulefill}\thinspace}

\end{document}